\documentclass{emulateapj}

\usepackage{lscape}
\def \kms{\ifmmode\,{\mbox{km}}\,{\mbox{s}}^{-1}\else km$\,$s$^{-1}$\fi}
\newcommand{\myemail}{lauren@physics.ubc.ca}
\newcommand{\taueff}{$\hat{\tau}$}
\newcommand{\dnf}{D$_{n}$(4000)}
\newcommand{\avgFe}{$\langle$Fe$\rangle$}
\newcommand{\etal}{et~al.\@}
\newcommand{\eg}{e.g.\@}
\newcommand{\ie}{i.e.\@}

\slugcomment{Submitted to ApJ.}

\shorttitle{Dust Sensitivity of Absorption-Line Indices}
\shortauthors{MacArthur}

\begin{document}

\title{Dust Sensitivity of Absorption-Line Indices}

\author{Lauren A. MacArthur}
\affil{Department of Physics \& Astronomy, University of British Columbia,
    6224 Agricultural Road, Vancouver, BC CANADA V6T 1Z1}
\email{\myemail}

\begin{abstract}
We investigate the effects of dust extinction on integrated absorption-line 
indices that are widely used to derive constraints on the ages and 
metallicities of composite stellar systems.  Typically, absorption-line 
studies have been performed on globular clusters or elliptical galaxies, 
which are mostly dust-free systems.  However, many recent studies of 
integrated stellar populations have focused on spiral galaxies which may 
contain significant amounts of dust. It is almost universally assumed
that the effects of dust extinction on absorption-line measurements are
entirely negligible given the narrow baseline of the spectral features, 
but no rigorous study has yet been performed to verify this conjecture.
In this analysis, we explore the sensitivity of the standard set of Lick
absorption-line indices, the higher-order Balmer line indices, the 
4000 \AA\ break, the near-IR calcium triplet indices, and the Rose indices 
to dust absorption according to population synthesis models that incorporate 
a multi-component model for the line and continuum attenuation due to dust. 
The latter takes into account the finite lifetime of stellar birth clouds.  
While dust does not greatly affect the line-index measurements for single 
stellar populations, its effect can be significant for the 4000 \AA\ break 
or when there is a significant amount of current star formation.
\end{abstract}


\section{Introduction}

Stellar population studies play a fundamental role in our understanding of
stellar evolution, initial mass functions (IMFs) associated with star cluster 
formation, and galaxy formation and evolution.  
With the aim of determining the star formation histories (SFHs) of stellar 
systems of all types (from star clusters to entire galaxies), many techniques 
have been developed, with varying success, to determine the 
luminosity-weighted ages and metallicities of nearby and distant stellar 
systems.  Important caveats hinder or limit the application of these techniques
(\eg\ Charlot, Worthey, \& Bressan 1996; MacArthur \etal\ 2004; 
Anders \etal\ 2004; Tantalo \& Chiosi 2004).
For example, the determination of ages and metallicities
of stellar populations has been plagued by the well-known age/metallicity 
degeneracy (Worthey 1994).  Access to both optical {\it and} infrared 
imaging partially lifts this degeneracy, but broad-band colors suffer 
further from a degeneracy due to dust reddening (\eg\ Bruzual, Magris, \& 
Calvet 1988; Witt, Thronson, \& Capuano 1992; de~Jong 1996; Bell \& de~Jong
2000; MacArthur \etal\ 2004),
which is particularly problematic in the context of gas rich, star forming 
systems.  Another method of breaking the age/metallicity degeneracy of 
stellar populations in unresolved systems involves 
measurement of surface brightness fluctuations (Worthey 1993; 
Liu, Charlot \& Graham 2000; 
Blakeslee, Vazdekis \& Ajhar 2001), which are most sensitive to contributions 
from the most luminous stars in the system -- typically evolved cool giants. 
Being sensitive to the second moment of the stellar luminosity function, 
surface brightness fluctuations
provide complementary information to the integrated colors (the first moment
of the stellar luminosity function).  This method, however, requires the 
assumption of a smooth underlying light distribution, and thus is generally 
only applicable to studies of nearby globular clusters, ellipticals, and 
spiral bulges.  This broad-band technique is also not applicable to dusty 
systems (\eg\ spiral galaxies), as the resulting clumpiness will also cause 
fluctuations in surface brightness and reddening will affect the 
fluctuation colors.
Turning to spectroscopy offers the possibility of overcoming the 
ambiguities of broad-band color-based analyses by studying the variation of
individual line strengths which, if defined over a sufficiently narrow 
wavelength range, should be insensitive to the effects of dust reddening.  
In this analysis we test this conjecture by exploring the effects of dust 
absorption on commonly-used spectroscopic age and metallicity indicators.

In the past three decades, tremendous progress has been made in the 
modeling of simple stellar populations (SSPs) (\eg\ Tinsley 1972; 
Bruzual \& Charlot 1993; Worthey 1994; Vazdekis 1999; Fioc \& 
Rocca-Volmerange 1997; 
Maraston 1998; Bruzual \& Charlot 2003). The current versions provide 
numerous observables such as high-resolution spectra, magnitudes, colors, 
mass-to-light ratios, and line-index measurements (including all those 
described above), for single bursts of star formation (SF) with a given 
IMF and metallicity at ages ranging from 0--20~Gyr.  In particular, the 
Bruzual \& Charlot (2003) models include a prescription for dust 
attenuation (described in \S\ref{sec:models}) that we adopt for this analysis.

Numerous studies of absorption-line indices in the integrated spectra of 
composite systems have sought to disentangle and constrain the ages and 
metallicities of their stellar populations (\eg\ Maraston \& Thomas 2000; 
Trager \etal\ 2000; Schiavon \etal\ 2002; Caldwell, Rose, \& Concannon 2003).  
The majority of these studies have focused on globular clusters (GCs) and 
elliptical galaxies, where dust is not conspicuous.  Recently, however, a 
number of studies have turned their attention to the stellar populations of 
spiral galaxies, where dust is an ineluctable hindrance 
(Fisher, Franx, \& Illingworth 1996; Goudfrooij, Gorgas, \& Jablonka 1999; 
Trager, Dalcanton, \& Weiner 1999; Proctor \& Samson 2002; 
Bergmann, J{\o}rgensen, \& Hill 2003; Kauffmann \etal\ 2003a; 
Falc{\' o}n-Barroso \etal\ 2003).
Reddening by dust has often been assumed to have a negligible effect on
on line indices, but no rigorous study has yet been performed to 
verify this conjecture.  This work presents the first such analysis.

The outline of the paper is as follows; a description of the line-indices
studied and their applicability as age and metallicity
discriminators is given in \S\ref{sec:indices}.  The stellar population 
synthesis models and dust prescription are described in \S\ref{sec:models}. 
In \S\ref{sec:results} we discuss the dust sensitivities of the Lick \& 
higher-order Balmer line indices, the 4000 \AA\ break, the \ion{Ca}{2} 
triplet indices, and the Rose indices, and the results are summarized in 
\S\ref{sec:disc}.  Finally, in Appendix~\ref{sec:fits}, we present age and 
metallicity fits to the model galaxies in several index--index planes and 
discuss the errors on the derived physical parameters due to dust extinction.

\section{Description of Indices Studied}\label{sec:indices}
\subsection{The Lick/IDS System}
The Lick/IDS system of 21 spectral line indices was designed to calibrate the 
strength of fundamental spectral features in stars and composite systems
(\eg\ Gorgas \etal\ 1993).  The indices measure the strength of a particular 
spectral feature (either atomic and defined as an equivalent width in \AA, or 
molecular and measured in magnitudes) 
relative to a pseudo-continuum on each side of the feature.  The most reliable 
indices have been calibrated as a function of stellar color (effective 
temperature), surface gravity, and metallicity (Gorgas \etal\ 1993; 
Worthey \etal\ 1994) allowing 
for the construction of semi-empirical population models (\eg\ Worthey 1994).  
The Lick indices are sensitive to the metallicity and 
age of stellar populations to varying degrees.  When compared with population 
models, diagnostic plots of age versus metallicity sensitive indices, such as
H$\beta$ versus Mg{\it b} or \avgFe, help break the age-metallicity 
degeneracy.  However, measurements of many of the Lick indices are quite 
sensitive to spectral resolution and, thus, to the velocity dispersion of the 
system (Gonz{\' a}lez 1993; Trager \etal\ 1998; Proctor \& Sansom 2002), and 
their use requires relatively high signal-to-noise data 
(S/N $\gtrsim$ 50/\AA; see Cardiel \etal\ 1998).  In addition, certain 
indices, H$\beta$ in particular, can suffer ``in-filling'' from nebular 
emission contamination, present even in early-type galaxies 
(Gonz{\' a}lez 1993; deZeeuw \etal\ 2002; Caldwell, Rose, \& Concannon 2003).  

\subsection{Higher Order Balmer Indices}
In order to overcome the problem of nebular emission fill in of the 
H$\beta$ feature, Worthey \& Ottaviani (1997; hereafter WO97) introduced two
pairs of indices that measure the higher-order Balmer lines H$\gamma$ and 
H$\delta$.  Two definitions for each feature were defined, the narrower 
version ($\Delta\lambda\simeq$ 20 \AA) is denoted with the subscript ``F'' 
(as it encompasses all of the Balmer line absorption from stars of spectral 
type F at the Lick/IDS resolution of 8--10 \AA), \eg\ H$\delta_{F}$, and the 
wider definition ($\Delta\lambda\simeq$ 40 \AA) has the subscript ``A'' (as 
it includes all of 
the absorption from A stars), \eg\ H$\gamma_{A}$.  The narrow indices
are more age-sensitive, but require higher S/N and resolution.
While their age-sensitivity is not as strong as for H$\beta$, the higher-order 
Balmer lines are much less affected by emission from ionized gases 
(\eg\ Osterbrock 1989).  Thus, when combined with a metallicity sensitive 
index, the WO97 indices provide a more reliable age estimate for star-forming 
galaxies.  Note, however, that the WO97 higher-order Balmer lines are poorly 
calibrated (Vazdekis 1999), likely due to the degrading 
resolution at the blue end of the IDS data. 

\subsection{The 4000 \AA\ Break}
Another widely used spectral index that obviates the need for high S/N and 
spectral resolution is the 4000 \AA\ break, a flux ratio that brackets the 
strongest discontinuity in the optical spectrum of a galaxy. 
The break arises due to the accumulation of a large number of spectral lines 
in a narrow wavelength region bluewards of 4000 \AA\ in stellar types cooler 
than G0 (Bruzual 1983; Gorgas \etal\ 1999). The main contribution to the 
$\la$ 4000 \AA\ opacity comes from atomic metals (\ion{Fe}{1}, \ion{Mg}{1}, 
\ion{Ca}{2}) and molecular CN, 
which decreases for hotter and more metal poor stars.  Thus the
4000 \AA\ break is weak for young and/or metal-poor stellar populations and 
strong for old, metal-rich galaxies (Kauffmann \etal\ 2003a).  In its 
original form, (Bruzual 1983), the 4000 \AA\ break, denoted
D(4000), was defined as the ratio of the average fluxes per frequency unit 
measured over the spectral ranges 4050--4250 \AA\ and 3750--3950 \AA. 
Cardiel \etal\ (1998) demonstrated that 
this discontinuity can be measured with a relative error of $\sim 10$\% 
with a S/N per \AA\ $\sim 1$, thus making it better suited for lower quality
data.  However, the D(4000) does have a few
drawbacks due to its long baseline.  These were partially alleviated with 
the introduction of a narrower definition (Balogh \etal\ 1999) denoted \dnf\ 
and measured over the ranges 4000--4100 \AA\ and 3850--3950 \AA. 
The narrow definition was designed to exploit two principal advantages: an 
improved agreement between multiple measurements of a given galaxy, and 
a weaker sensitivity to reddening effects.  However, using 
their narrow \dnf\ index, Balogh \etal\ (1999) still had to invoke dust 
reddening as a cause for the large \dnf\ values in a number of their 
$z\sim 0.3$ galaxies, \ie\ \dnf\ is not impervious to dust effects.  
Nevertheless,
often found in the literature is the statement that the \dnf\ is 
{\it insensitive} to dust attenuation effects.  For example, Kauffmann \etal\ 
(2003b) use the amplitude of the \dnf\ in combination with the strength of the 
H$\delta_{A}$ index of WO97 as diagnostics for the SFH of the host galaxies, 
from which they 
infer the dust attenuation by comparing observed to model colors, a method 
that relies heavily on the assumption of dust insensitivity of the \dnf\ and 
H$\delta_{A}$ indices.

\subsection{The Near-IR \ion{Ca}{2} Triplet Indices}
There has been a great effort recently to extend stellar population 
studies to the near-IR region, focusing on the \ion{Ca}{2}
triplet as one of the most prominent features in the near-IR spectrum of
cool stars (from spectral types of about F5 to M2) 
(Cenarro \etal\ 2001a,b; Vazdekis \etal\ 2003).
Three ``Lick-style'' indices designed to measure the strengths of the
\ion{Ca}{2} triplet lines ($\lambda\lambda$8498.02, 8542.09, 8662.14 \AA), and
denoted Ca1, Ca2, and Ca3, have been defined
and redefined by several authors (\eg\ Jones, Alloin, \& Jones 1984; 
Armandroff \& Zinn 1988; Diaz, Terlevich, \& Terlevich 1989, hereafter DTT; 
Delisle \& Hardy 1992).  However, measurement of a reliable continuum for
these indices is difficult due to the strong and crowded absorption
features in the vicinity of the \ion{Ca}{2} lines (largely from
\ion{Fe}{1}, \ion{Mg}{1}, and TiO), as well as
significant blending with the hydrogen Paschen series 
whose absorption is present in stars of types G3 and hotter. 

To overcome these problems, Cenarro \etal\ (2001a) defined a new set of 
\ion{Ca}{2} triplet indices specifically
designed for measurement in integrated galactic spectra.  The new definitions
are categorized as ``generic'' indices, which have advantages over the
classical ``Lick-style'' indices when looking at adjacent absorption lines in
regions where the continuum is crowded with spectral features.  Their CaT
index includes the strenghts of all three \ion{Ca}{2} lines
and uses a combination of 5 continuum bandpasses (see Cenarro \etal\ 2001a for
details on the measurement of generic indices).  Additionally, they 
define the generic index PaT, which measures the strength of three of the 
H Paschen series lines that are free from Ca contamination.  Finally, the 
index CaT* is designed to remove the H Paschen line contamination from CaT
making it a reliable indicator of the pure \ion{Ca}{2} triplet strength. 
The index is given by CaT* = CaT - 0.93 PaT.  Vazdekis \etal\ (2003) discuss 
the behavior of these features for SSPs as predicted by recent evolutionary 
synthesis models.  In the current analysis we also explore the dust 
sensitivity of the three classical \ion{Ca}{2} triplet indices, as defined 
in DTT, and the generic indices of Cenarro \etal\ (2001a).

\subsection{The Rose Indices}
The optical Rose (1984, 1985) indices were developed in part to overcome the 
difficulties in identifying the continuum in crowded spectral regions.
The Rose indices maximize the contribution of a given feature by using a 
measurement of the ratio of its central line intensity to that of a close 
reference line, without recourse to the (pseudo-) continuum level.  
Information about equivalent widths is however lost in such relative 
measurements, and the Rose indices are not ideal for separating age and 
metallicity effects.  Nonetheless, some of the Rose indices are quite 
sensitive to metallicity (see Vazdekis 1999) and, in combination with a
sensitive age indicator (\eg\ H$\beta$), could help disentangle the two 
effects.  The Rose indices also provide a unique and sensitive test for the 
presence of early-type stars (\ie\ a post-starburst, and potentially dusty, 
population) in an integrated spectrum as well as a probe of the relative 
contribution of dwarf and evolved (red giant branch) stars to the integrated 
spectrum of a galaxy (Caldwell, Rose, \& Concannon 2003). 
As with the Lick indices, it 
is usually assumed that the Rose indices are insensitive to dust reddening 
(\eg\ Leonardi \& Worthey 2000), an assumption we examine in this study.

Finally, we also investigate dust effects on the pseudo-equivalent width,
``Lick-style'', indices of Rose (1994) and Jones \& Worthey (1995) which have 
extremely narrow definitions (and are denoted with the subscript ``HR'').  The 
H$\gamma_{{\mbox{\scriptsize HR}}}$ index has been identified as one of the 
most sensitive age distriminators (Jones \& Worthey 1995), but these narrow 
features suffer a worse resolution sensitivity than the WO97 indices and thus
can only be used with high quality data of low velocity-dispersion systems.
Vazdekis \& Arimoto (1999) and Vazdekis \etal\ (2001) confronted 
the velocity dispersion sensitivity of these narrow indices by 
defining a set of four H$\gamma$ indices that take resolution effects into 
account, allowing for reliable measurements in galaxies with velocity 
dispersions up to $\sigma\sim 300$ \kms.  Note, however, that the Vazdekis \& 
Arimoto (1999) indices require S/N $>$ 200--400/\AA\ and have strong error 
covariances due to overlapping pseudo-continuum and central bandpasses, making
them much harder to measure than the WO97 indices.  The response of these 
indices to dust extinction is similar to the WO97 H$\gamma_{F}$ and the 
Jones \& Worthey (1995) H$\gamma_{{\mbox{\scriptsize HR}}}$ indices, so the results are 
not shown.

\section{Models}\label{sec:models}

The stellar population synthesis models used for this analysis are those of 
Bruzual \& Charlot (2003; hereafter GALAXEV).  The high resolution models 
using the Padova evolutionary tracks (Bertelli \etal\ 1994) and the 
Chabrier (2003) IMF were adopted.  These models include SSP spectra in the 
wavelength range 3200--9500 \AA\ with a resolution of 3 \AA, metallicities 
ranging from $Z =$ 0.0001--0.05 (or 0.005--2.5 times Z$_{\odot}$), and ages 
ranging from 0--20~Gyr (in 220 unequally-spaced time steps).  For each time 
step, a number of integrated quantities 
are provided such as magnitudes and colors in many different filter systems,
line-index strengths for all definitions in the Lick/IDS system, WO97, the 
two definitions of the 4000\AA\ break, and the DTT \ion{Ca}{2} indices.  
The line strengths used here are computed directly from the high-resolution 
model spectra, \ie\ the spectra have not been transformed to the Lick/IDS
system (see \eg\ WO97).  The Rose, \ion{Ca}{2}, and ``HR'' spectral indices, 
which are not provided with the BC03 model outputs, were computed with a 
public \texttt{fortran} program by 
A. Vazdekis\footnote{See http://www.iac.es/galeria/vazdekis/}.

\begin{figure*}  
\begin{center}
\includegraphics[width=4.5in]{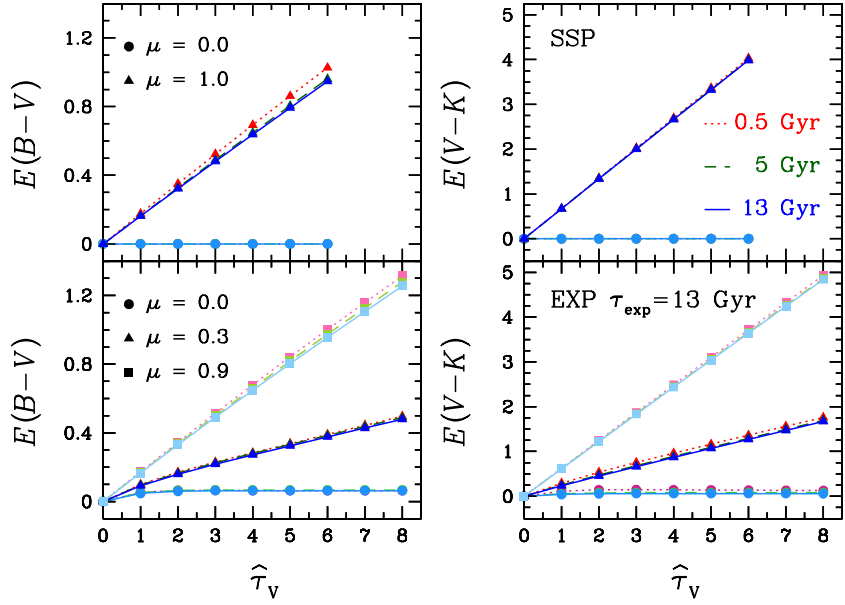}
\vspace{-1.0in}
\caption{Color excesses $E(B-V)$ [left panels] and $E(V-K)$ [right panels]
         as a function of \taueff$_V$ for solar metallicity SSPs with
         $\mu$ = 0.0 (circles) \& 1.0 (triangles) [top panels] and an 
         exponential SFH with $\tau_{exp} = 13$~Gyr for $\mu$ = 0.0 (circles), 
         0.3 (triangles), and 0.9 (squares) [bottom panels].  Different 
         ages are represented by: 0.5 (dotted lines; red shades), 
         5 (dashed lines; green shades), 
          \& 13 Gyr (solid lines; blue shades).
         \label{fig:EBmV}}
\end{center}
\end{figure*}

The GALAXEV distribution 
allows for the computation of attenuation effects due to dust according to 
the two-component model of Charlot \& Fall (2000; hereafter CF00).  The two 
adjustable parameters of this model are \taueff$_{V}$, the total 
effective $V$-band optical depth affecting stars younger than 10${^7}$ yr, 
and $\mu$, the fraction of the total dust absorption contributed by diffuse 
interstellar medium (cirrus) dust.  That is,
\begin{equation}
\hat{\tau}_{\lambda} = \left\{ \begin{array}{lll}
\hat{\tau}_{V}({\lambda}/5500 \,\,{\mbox{\AA}})^{-0.7} &  {\mbox{for}} & t \le 10^{7}\,\,{\mbox{yr},} \\
\mu\,\hat{\tau}_{V}({\lambda}/5500 \,\,{\mbox{\AA}})^{-0.7} &   {\mbox{for}} & t > 10^{7}\,\,{\mbox{yr},}
\end{array}
\right.
\label{eq:dust}
\end{equation}
where $t$ is the age of any single stellar generation.  This model has the
singular feature of accounting 
for the finite lifetime ($\sim 10^7$ yr) of stellar 
birth clouds\footnote{Introduced to resolve the apparent conflict between the 
attenuation of line and continuum photons in their sample of starburst 
galaxies.}.  The 
wavelength dependence of the effective absorption curve (proportional 
to $\lambda^{-0.7}$) was constrained to reproduce the observed relation 
between the ratio of far-infrared to ultraviolet luminosities, and the 
ultraviolet spectral slope of their starburst galaxies (see also the GRASIL 
models of Silva \etal\ 1998). 
Note that \taueff$_{\lambda}$ on the left-hand side of 
equation~(\ref{eq:dust}) is a function of time and $\mu$.  
As a result, even if $\mu=0$ (no cirrus dust component), in the presence of
young stellar populations ($t \le 10^{7}$ yr), dust will still contribute
some extinction (see Fig.~\ref{fig:EBmV}).
Note that in the adopted dust model of CF00, the model spectra are 
representative of the total integrated light of a galaxy, \ie, neither radial 
nor inclination dependent information is available.

Other prescriptions for dust attenuation in galaxies exist (\eg\ Witt, 
Thronson, \& Capuano 1992; Byun, Freeman, \& Kylafis 1994; de~Jong 1996; 
Gordon \etal\ 2001), a few of which have been coupled with spectral synthesis 
and photo-ionization codes. For example, Moy \etal\ (2001) interfaced the 
P\'{E}GASE population synthesis models of Fioc \& Rocca-Volmerange (1997) with 
the CLOUDY photo-ionization code of Ferland (2002).  These models include a 
treatment for dust attenuation processes, but do not consider nebular 
emission.  They also adopt a simple screen approximation for the dust
distribution, an unrealistic geometry for galaxies.
Similarly, Panuzzo \etal\ (2003) coupled the same CLOUDY code with their 
spectrophotometric synthesis model GRASIL (Silva \etal\ 1998), to yield
a complete treatment of dust reprocessing with more realistic geometries and
account for the age dependence of molecular birth clouds (as in CF00).  While
these models may be more realistic and provide directional information that
is lacking in the CF00 models, the resolution of the output from these models
is not sufficient for accurate line-index measurements.  Future implementations
of these models may allow for a comparative analysis, but
at present, the CF00 dust models, coupled with the BC03 SPS models, provide
(to the best of our knowledge) the only suitable output for the current study. 

In a study of 705 non-Seyfert galaxies drawn from the Stromlo-APM redshift 
survey (Loveday \etal\ 1996), Charlot \etal\ (2002) determined parameters for 
the CF00 models that reproduce the observed integrated spectral properties of 
the nearby star-forming galaxies. These cover the ranges:
$0.2 \le Z/Z_{\odot} \le 4.0$,
$0.01 \le$ \taueff$_{V} \le 4.0$, and
$0.2 \le \mu \le 1.0$,
for constant and exponential SFHs with ages 
$10^{7}$ yr $\le t \le 10^{10}$ yr.
Two time-scales were adopted for the exponentially declining star 
formation rate: $\tau_{exp} = 0.1$ and 6.0 Gyr.  Our model realizations
cover roughly the same parameter space.  For the SSP models 
we consider values of $\mu =$ 0 and 1 only since, for the
SSP ages shown ($t > 10^{7}$ yr), different values of $\mu$ are equivalent to 
different values of \taueff$_{V}$.  For the exponential SFHs, for which there 
are always populations with $t \le 10^{7}$ yr present, we present models 
with $0 < \mu < 1$ and extend to \taueff$_{V}$ = 8 
(so for the $\mu=0.3$ case, the plots effectively extend to \taueff$_{V}=2.4$).
To express the effective dust reddening in a more familiar form, in 
Figure~\ref{fig:EBmV} we plot the color excesses $E(B-V)$ and $E(V-K)$ as a 
function of \taueff$_V$ resulting from the dust models used in this analysis.
The upper panels are for solar metallicity SSPs with $\mu$ = 0.0 (circles) \&
1.0 (triangles) and the lower panels show exponential SFH with 
$\tau_{exp} = 13$~Gyr for $\mu$ = 0.0 (circles), 0.3 (triangles), \&
0.9 (squares) at ages of 0.5 (dotted lines; red shades),  5 (dashed lines; 
green shades), \& 13 Gyr (solid lines; blue shades).

\begin{figure*} 
\begin{center}
\vspace{- 0.45in}
\includegraphics[width=5.0in]{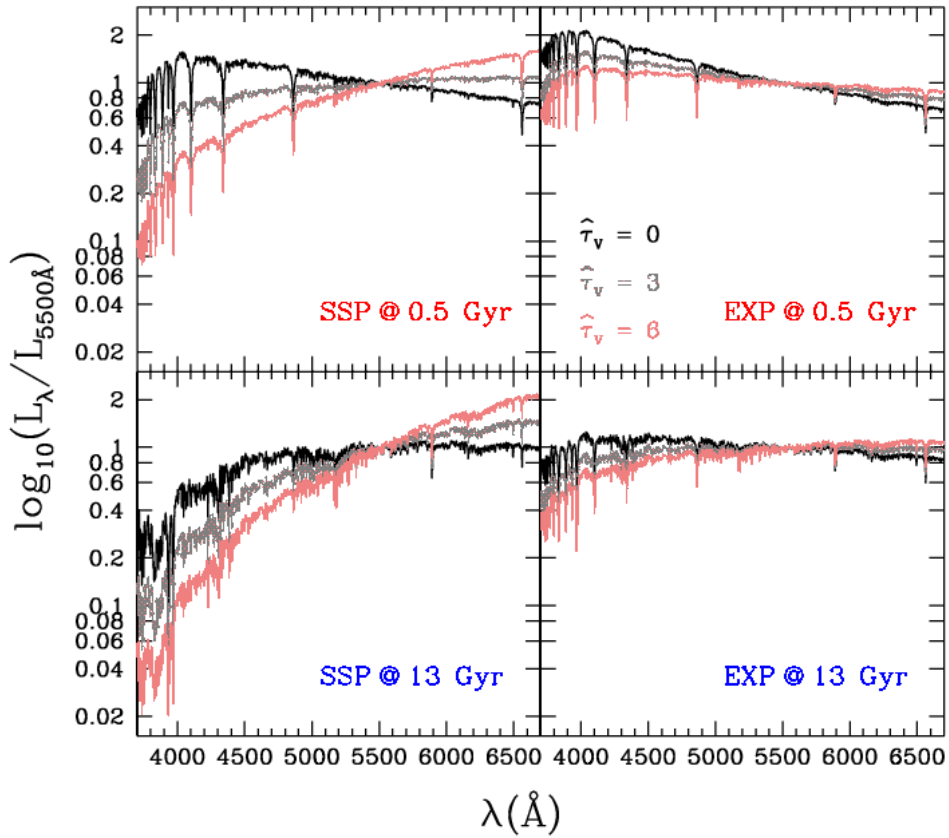}
\vspace{- 0.1in}
\caption{Comparison of solar metallicity SEDs with \taueff$_V$ = 0 (black), 
         3 (gray), \& 6 (pink) 
         at ages of 0.5 [top panels] \& 13 Gyr [bottom panels] for 
         SSP models with $\mu$ = 1.0 [left panels] and exponential SFH 
         models with $\tau_{exp} = 13$~Gyr and $\mu$ = 0.3 [right panels].
         \label{fig:SED_both}}
\end{center}
\end{figure*}

Before examining the response of the individual indices to the dust 
reddening, it is useful to look at the spectral energy distribution (SED)
of the models.  Figure~\ref{fig:SED_both} shows the model SEDs in the 
3700--6700 \AA\ range for SSPs [left panels] and exponential SFHs with 
$\tau_{exp}=13$~Gyr [right panels].  All spectra have been normalized to 
5500 \AA.
The SSP SEDs look entirely as expected, with the 0.5~Gyr 
population [top right panel] closely resembling an early-type (A--F) stellar 
spectrum with a blue continuum, strong Balmer features, but few metallic 
features.  The 13~Gyr spectrum [bottom right panel] more closely resembles a 
later-type (G--K) stellar spectrum with a redder continuum, weaker Balmer-line 
strengths, and more metallic absorption features.  In both cases, the dust 
extincted profiles exhibit the overall low frequency reddening of the SED due 
to the $\lambda^{-0.7}$ dependence of the dust model.  On the other hand, the 
unreddened 0.5~Gyr SED for an exponential SFH with $\tau_{exp}=13$~Gyr 
(Fig.~\ref{fig:SED_both} [top right panel, black curve]) is even bluer than 
its SSP counterpart because of the presence of young stars ($<0.5$~Gyr) from 
the ongoing SF.  The contrast between the SSP and exponential SFHs is even
more striking at 13~Gyr.  The exponential SFH SED (Fig.~\ref{fig:SED_both}
[bottom right panel, black curve]) is much bluer than its SSP counterpart
[bottom left panel] due to the ongoing SF, but it also shows metallic features 
due to the presence of 
older stars ($\le 13$~Gyr).  In the exponential SFH case, the dust reddening
is more complicated due to the presence of young stars that still live in
their birth clouds ($< 10^{7}$~yr).  These stars are more extincted than
the older stars, the cirrus extinction component, when $\mu<1.0$, but being
intrinsically bright, still contribute significantly to the total flux.
In Figure~\ref{fig:SED_both} [right panels] we show the case for $\mu=0.3$.  
Clearly, any measurement made over a long baseline (\eg\ colors) will be 
affected by the dust reddening.  However, gauging its effects on 
absorption-line indices from a cursory examination of the SEDs is not a 
straightforward task.  In the next section we examine individual 
features and compare the dust-free indices with those computed with dust.

\section{Results}\label{sec:results}

\subsection{Dust Sensitivity of the Lick Indices}\label{sec:lick}

\begin{figure*} 
\begin{center}
\vspace{- 0.5in}
\includegraphics[width=5.0in]{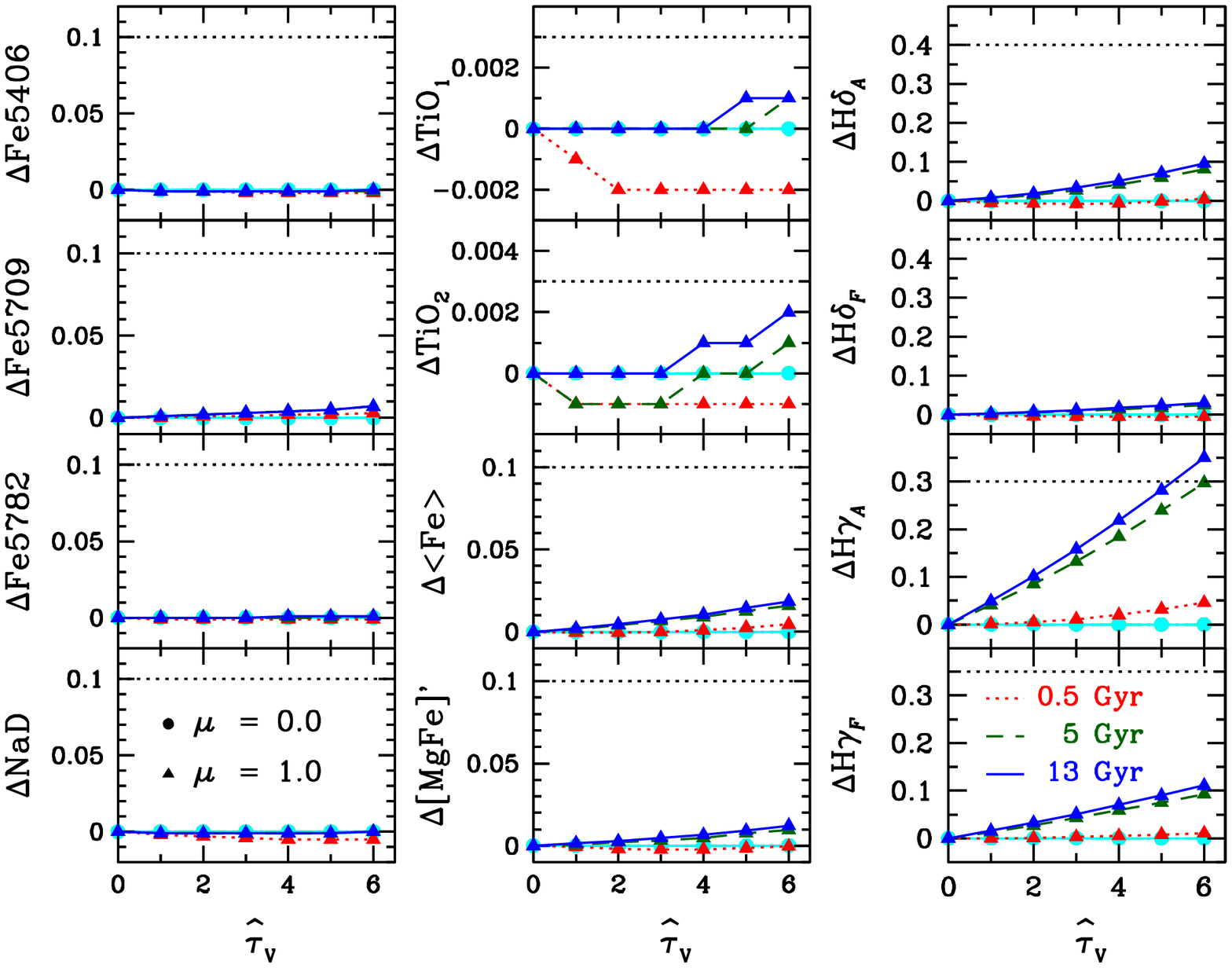}
\vspace{- 1.0in}
\caption{Lick and WO97 index differences, $\Delta{index} \equiv 
         index$(\taueff$_{V}) - index($\taueff$_{V} = 0)$, as a function of 
         \taueff$_V$ for solar metallicity SSPs.  Different 
         ages are represented by: 0.5 (dotted lines; red shades), 
         5 (dashed lines; green shades),  
         \& 13 Gyr (solid lines; blue shades).
         Different values of $\mu$ are represented as 0.0 (circles), 
         and 1.0 (triangles). The black horizontal dotted lines 
         represent typical measurement errors for the different indices.
         \label{fig:SSPs}}
\end{center}
\end{figure*}

\begin{figure*}
\begin{center}
\vspace{- 0.6in}
\includegraphics[width=5.0in]{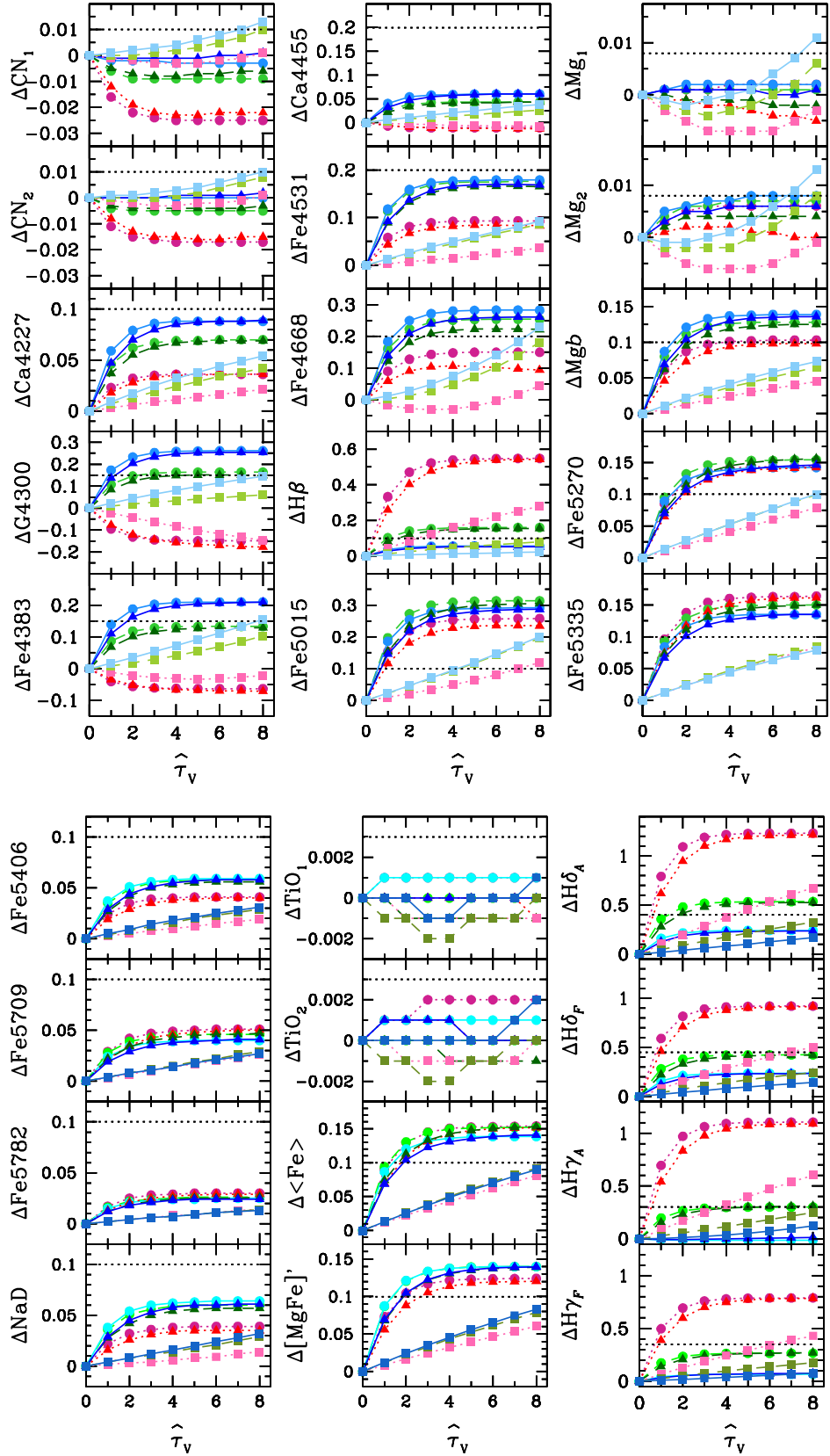} 
\vspace{- 1.0in}
\caption{Same as Figure~\ref{fig:SSPs} but for 
         for an exponential SFH with $\tau_{exp} = 13$ Gyr and 
         values of $\mu$ are represented as 0.0 (circles), 0.3 (triangles),
         and 0.9 (squares).\label{fig:expsfh}}
\end{center}
\end{figure*}

The response of the Lick indices as a function of \taueff$_{V}$, $\mu$, 
and age for solar metallicity ($Z_{\odot}=0.02$) and abundance ratio 
([$\alpha$/Fe] = 0) SSP models is shown in Figure~\ref{fig:SSPs}.  In each 
panel, we plot $\Delta{index}$ versus \taueff$_{V}$, where $\Delta{index}$ is 
the difference between the indices measured with and without dust (for the 
same age \& SFH), 
\ie\ $\Delta{index} \equiv index$(\taueff$_{V}) - index($\taueff$_{V} = 0)$.  
Results are shown for model ages of 0.5 (dotted lines; red shades),  
5 (dashed lines; green shades), \& 13 Gyr (solid lines; blue shades).  
The two values for $\mu$ of 0.0 \& 1.0 are 
denoted by circles and triangles, respectively.
The horizontal dotted lines in Figures~\ref{fig:SSPs}, \ref{fig:expsfh}, 
\& \ref{fig:d4000} -- \ref{fig:Rose_expsfh} denote 
typical measurement errors for the 
different indices; $\sim 0.1$~\AA\ for the atomic Lick indices, 
0.01 mag for the molecular Lick indices, 0.4~\AA\ for the WO97 indices, 
0.03 for D(4000) and \dnf, and  $\sim$0.02--0.05 for the Rose indices
(Gorgas \etal\ 1999; Jones \& Worthey 1995; Kauffmann \etal\ 2003a; 
Falc{\' o}n-Barroso \etal\ 2003; Caldwell, Rose, \& Concannon 2003;
J.J. Gonz{\' a}lez 2004, private comm.), and serve as a guide for the 
magnitude of the effect from dust required to be detectable above the noise.  
All model realizations presented here have solar metallicity, 
$Z_{\odot} = 0.02$.  We have also considered the other metallicities
in the GALAXEV models.  Generally, the dust effects are slightly more 
conspicuous at $Z=0.05$, but diminish with decreasing metallicity of the 
stellar population.
\begin{figure*} 
\begin{center}
\includegraphics[width=5.0in]{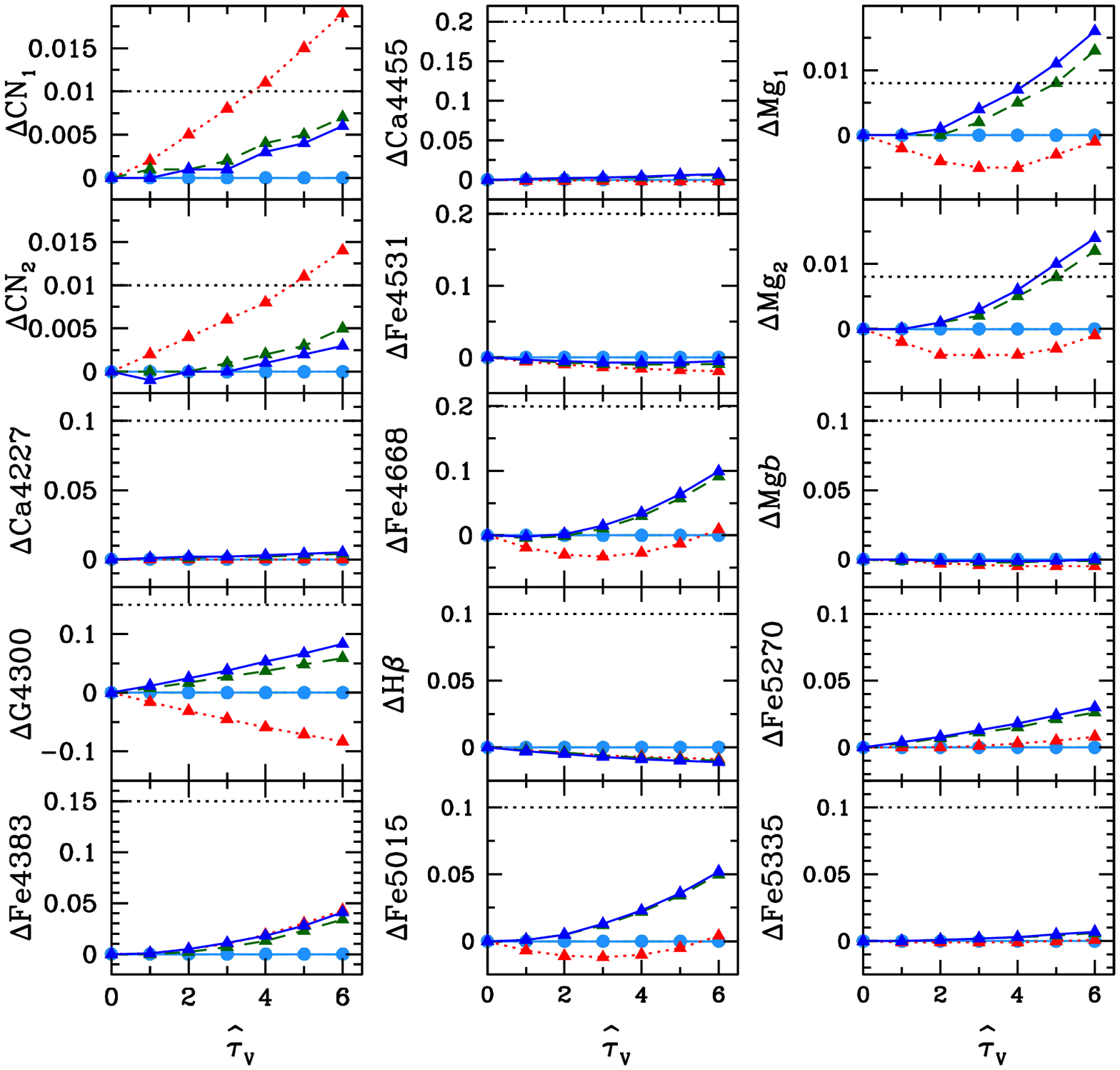}
\end{center}
\end{figure*}

In most cases, the index measurements do not deviate considerably from 
their dust-free values in the SSP models, as expected.  Since we consider
SSPs at ages of 0.5 Gyr and greater, by definition, models with $\mu = 0$ 
have no dust extinction (thus circle symbols lie at 0 for all values
of \taueff$_{V}$ in Fig.~\ref{fig:SSPs}).  For $\mu = 1$ models,
deviations from the dust-free index measurements do occur.  There are a few 
cases of note that show significant sensitivity to the dust extinction, such
as the molecular indices Mg$_1$ \& Mg$_2$, and CN$_1$ \& CN$_2$.  This 
is not surprising since these indices have the longest wavelength baseline 
($\Delta\lambda\sim$400 \& 200 \AA\ respectively).  As evidenced by the 
molecular Mg indices, index measurements in the presence of dust can be 
non-linear and age dependent, thus a simple ``dust-correction'' poses a 
formidable
challenge.  The effect of extinction on the Mg indices goes in the 
opposite direction for young ages as it does for older ages (compare dotted
and dashed/solid lines in Fig.~\ref{fig:SSPs}).  
Also, the CN indices are more severely affected at younger ages whereas 
the Mg indices are more affected at young ages for small \taueff$_{V} \la 3$, 
but at older ages for larger \taueff$_{V} > 4$. 

Of the Lick atomic indices, those showing the greatest sensitivity to dust 
are Fe5015 \& Fe4668, both having the longest wavelength baselines of all 
atomic indices, but their deviations from the dust-free case are always
smaller than the typical measurement errors.
In addition to the standard Lick indices, we also show the combination
indices, \avgFe = (Fe5270+Fe5335)/2 and [MgFe], first 
introduced by Gonz{\' a}lez (1993).  For the latter we use the slightly 
modified definition of Thomas, Maraston, \& Bender (2003), 
[MgFe]$\arcmin$ = [Mg{\it b}(0.72$*$Fe5270+0.28$*$Fe5335)]$^{1/2}$, which
has the advantage of being completely independent of the element abundance 
ratio ([$\alpha$/Fe]).  These combination indices vary very little with 
\taueff${_V}$ for the SSP models (Fig.~\ref{fig:SSPs}).

The higher-order Balmer indices, shown in Figure~\ref{fig:SSPs}, also show 
some sensitivity to dust at older ages in the SSP models, particularly the 
H$\gamma_{A}$ index.  Measurement errors for these indices are quite large 
compared to the other Lick indices (due to their narrow baselines), thus 
any reddening effect would likely not be detectable.

Unlike GCs and elliptical galaxies, the assumption of SSP-like star formation 
is clearly not applicable for spiral galaxies.
While the true SFHs of the latter are poorly known in detail, 
they are often approximated with an exponentially declining SFR, parameterized
by a star formation timescale $\tau_{exp}$ (\eg\ Bell \& de~Jong 2000;
MacArthur \etal\ 2004).  In Figure~\ref{fig:expsfh} we show the index responses
to dust for an exponential SFH with $\tau_{exp}=13$~Gyr, which is reasonably 
close to a constant SF rate so there is a significant amount of ``current'' 
SF at each of the three epochs considered.
\begin{figure*}
\begin{center}
\includegraphics[width=5.0in]{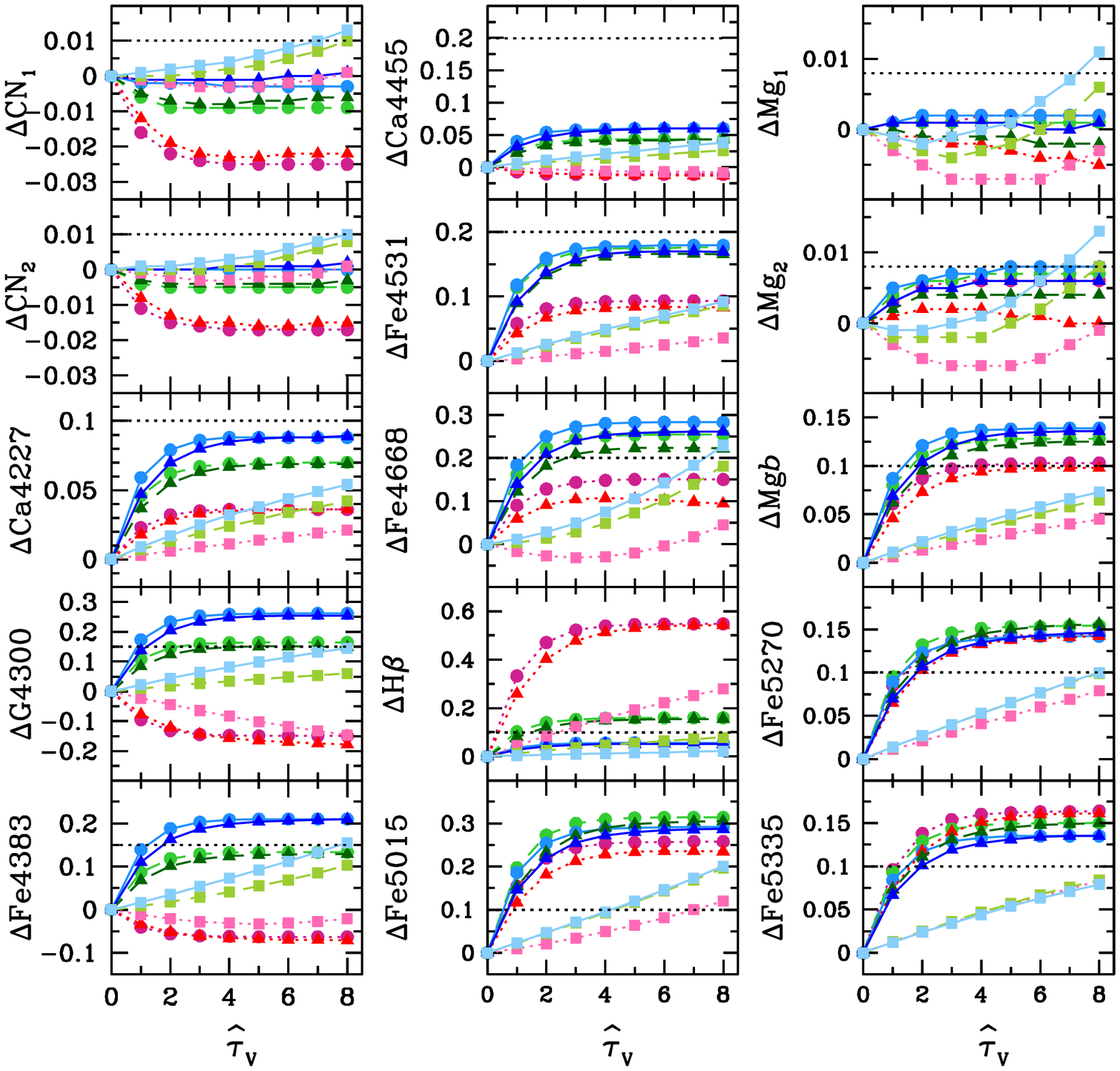} 
\end{center}
\end{figure*}

Models with current SF yield considerably different results than the SSPs, as 
seen in Figure~\ref{fig:expsfh}.  Almost all of the Lick indices
show sensitivity to dust, some more so at young ages (\eg\ CN$_1$, 
CN$_2$, H$\beta$, and all four WO97 indices), others more so
at older ages (\eg\ Ca4227, Fe4531, Fe4668, Mg{\it b}, and NaD),
and some show differences of equal magnitude, but in opposite directions
at old and young ages (\eg\ G4300, Fe4383, and the molecular Mg indices).  
The situation is
undoubtedly complex and can be attributed to the fact that, with current SF
occurring at all epochs, there is a significant and non-linear 
amount of extinction of the hottest, youngest stars which 
contribute considerably to the total optical flux. 
As such, dust effects are often stronger for models with smaller $\mu$ 
and these effects saturate at optical depths above about \taueff$_{V}=3$ 
(at which the birth clouds become completely optically thick).

The situation for mixed stellar populations with significant amounts of 
current SF is clearly complicated and not easily prescribed.  However, for
many of the indices (\eg\ Ca4227, Ca4455, Fe5406, Fe5709, Fe5782, NaD, 
TiO$_1$, and TiO$_2$), the magnitude of the dust effects is never close to 
that of current typical measurement errors, and thus would not be detected.  
Unfortunately, however, the most often used indices (\eg\ H$\beta$, 
Mg{\it b}, the composite indices \avgFe\ and [MgFe]$\arcmin$, and the 
WO97 indices) are also the most seriously affected.

\begin{figure*}
\begin{center}
\includegraphics[width=4.7in]{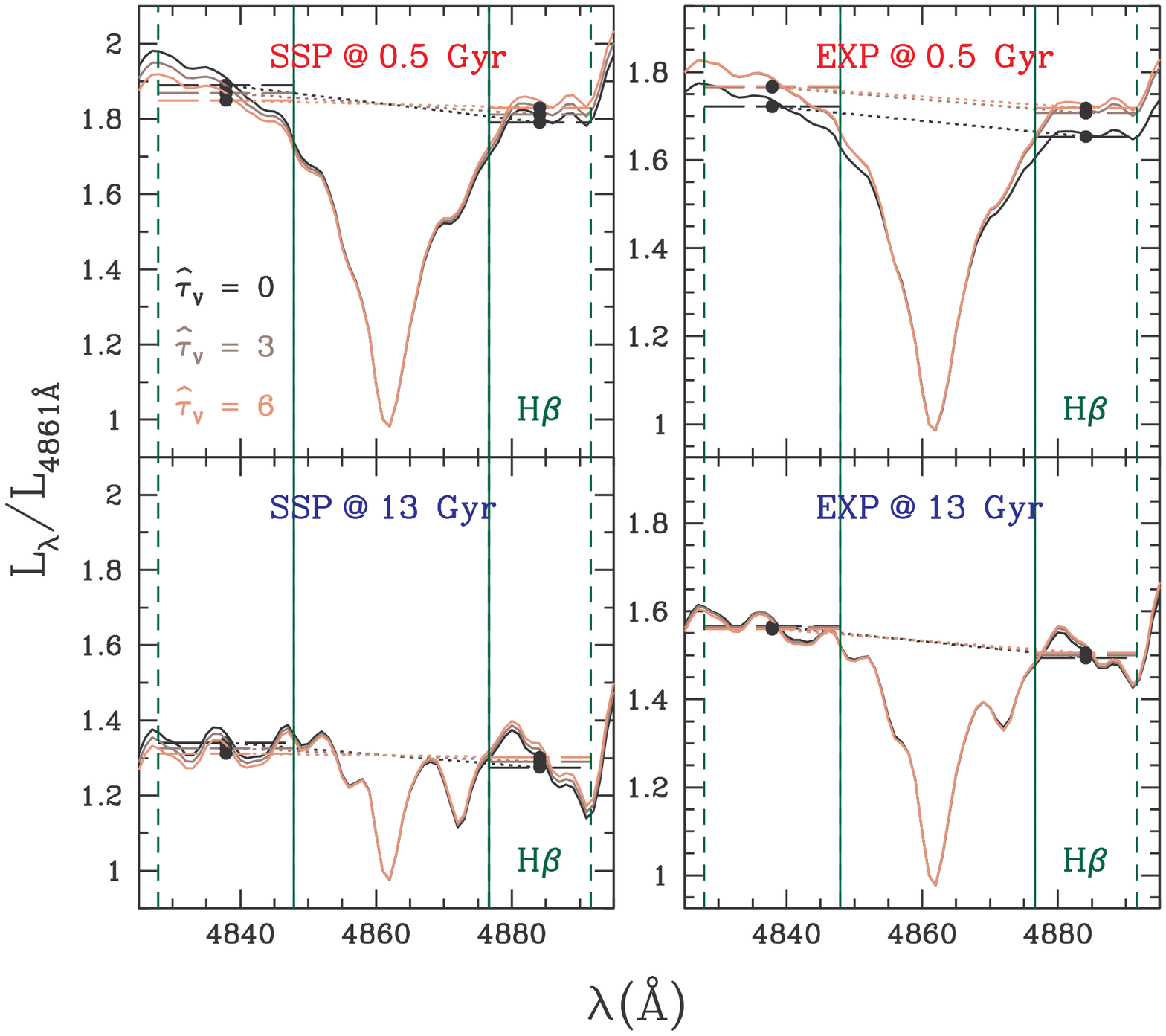} 
\vspace{- 0.55in}
\caption{Comparison of the H$\beta$ index for solar metallicity   
         SSP models with  $\mu$ = 1.0 [left panels], and  
         exponential SFH  $\tau_{exp} = 13$~Gyr models with $\mu$ = 0.3.
         [left panels].  All panels show models with \taueff$_V$ = 0 (black), 
         3 (gray), \& 6 (pink) at ages of 0.5 [top panels] 
         \& 13 Gyr [bottom panels].  The index red and blue pseudo-continuum 
         band limits are marked by the green dashed vertical lines and the 
         central band by the solid vertical lines.  The 
         dotted lines mark the pseudo-continuum level for each spectrum.  
         The spectra are all normalized to their flux at 4861 \AA\ (the 
         center of the H$\beta$ line).
         \label{fig:Hbeta_both}}
\end{center}
\end{figure*}

It is useful to take a closer look at the cause for the erratic behavior
in some of the indices in response to the dust extinction.  
In Figure~\ref{fig:Hbeta_both} 
we focus on the H$\beta$ index. Plotted in each panel are
solar metallicity SEDs for \taueff$_V$ = 0 (black), 3 (gray), \& 6 (pink)
at ages of 0.5 [top panels], \& 13 Gyr [bottom panels].  The index
red and blue pseudo-continuum band limits are marked by the green dashed
vertical lines and the central band by the solid vertical lines.  The 
dotted lines mark the pseudo-continuum level for each spectrum.  
\begin{figure*}
\begin{center}
\vspace{- 0.24in}
\includegraphics[width=4.65in]{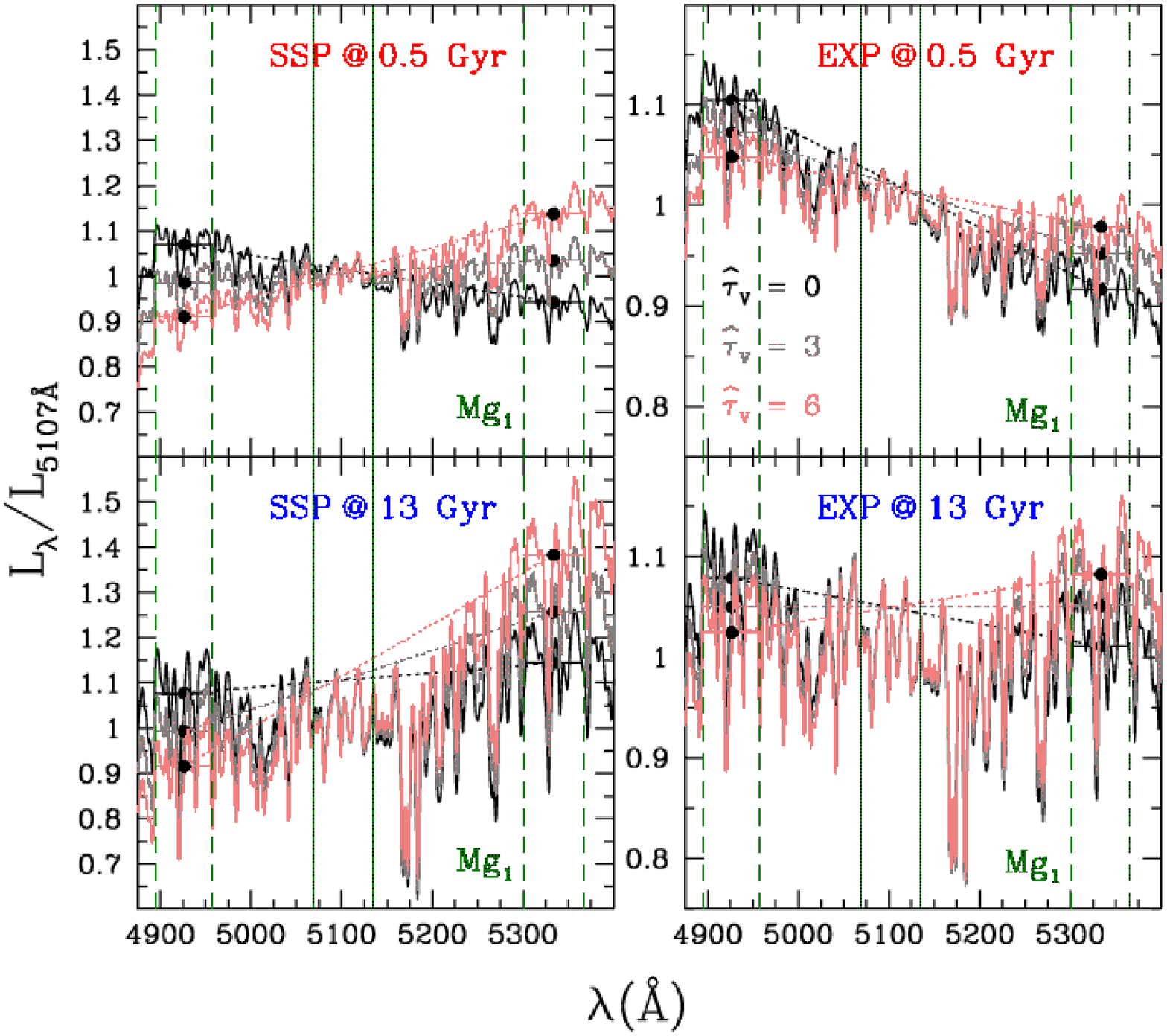} 
\vspace{- 0.15in}
\caption{Same as Figure~\ref{fig:Hbeta_both} but for Mg$_1$.
         The spectra are all normalized to their flux at 5107 \AA\ 
         (roughly the center of the index central passband).
         \label{fig:Mg1_both}}
\end{center}
\end{figure*}
The spectra are all normalized to their 4861 \AA\ flux (H$\beta$ line center). 
Figure~\ref{fig:Hbeta_both} plots SSPs [right panels] with $\mu$ = 1.0 
exponential SFHs [left panels] with $\mu$ = 0.3.  There is very 
little effect in the SSP index measurement, the change in slope of the SED
is matched by the change in slope of the
pseudo-continuum.  However, for the exponential SFH at 0.5 Gyr 
(Fig.~\ref{fig:Hbeta_both} [top right panel]), the continuum
is increased relative to the depth of the absorption-line in the dust
extincted models.  This can be understood by recognizing that the young OB 
stars that suffer significant extinction have lower H$\beta$ values.  Thus,
by hiding these low H$\beta$ index stars, the index is effectively increased.
In fact, this is the case for all of the Balmer lines, hence all of the 
Balmer-line indices show similar behavior.  This also highlights a major 
limitation when using Balmer-line indices in determining young ages 
($\la 0.5$ Gyr).  The models become degenerate in age here because there are
two possible ages that can be inferred at these Balmer index values, \ie\
the Balmer indices are double-valued in the region of young stellar population
ages.  This point and its resulting limitations are further emphasized
in Appendix~\ref{sec:fits}.

For the molecular indices, which have much broader baselines and crowding of 
spectral lines in their passband regions, the situation can be more 
complicated.  For instance, in Figure~\ref{fig:Mg1_both} 
we focus on the Mg$_1$ index.  The color codes, line-types, and SFHs are the 
same as in Figure~\ref{fig:Hbeta_both}.
In this case, given the much broader baseline, the slope
of the pseudo-continuum in the extincted models changes quite dramatically.
In the 0.5 Gyr SSP (Fig.~\ref{fig:Mg1_both} [top left panel]) and 13 Gyr 
exponential SFH (Fig.~\ref{fig:Mg1_both} [bottom right panel]) cases it 
changes from being a negative slope in the 
dust-free model (black) to positive in the most extincted curve (pink).  These
pseudo-continua pass through a crowded set of absorption lines in the central
passband, and a simple inspection of these figures does not reveal 
which model would yield larger index measurements.  It is clear from these 
figures that a simple dust correction prescription for all of the indices is 
cannot be achieved.

\subsection{Dust sensitivity of the 4000\AA\ Break} \label{sec:d4000}

\begin{figure*}
\begin{center}
\includegraphics[width=5.0in]{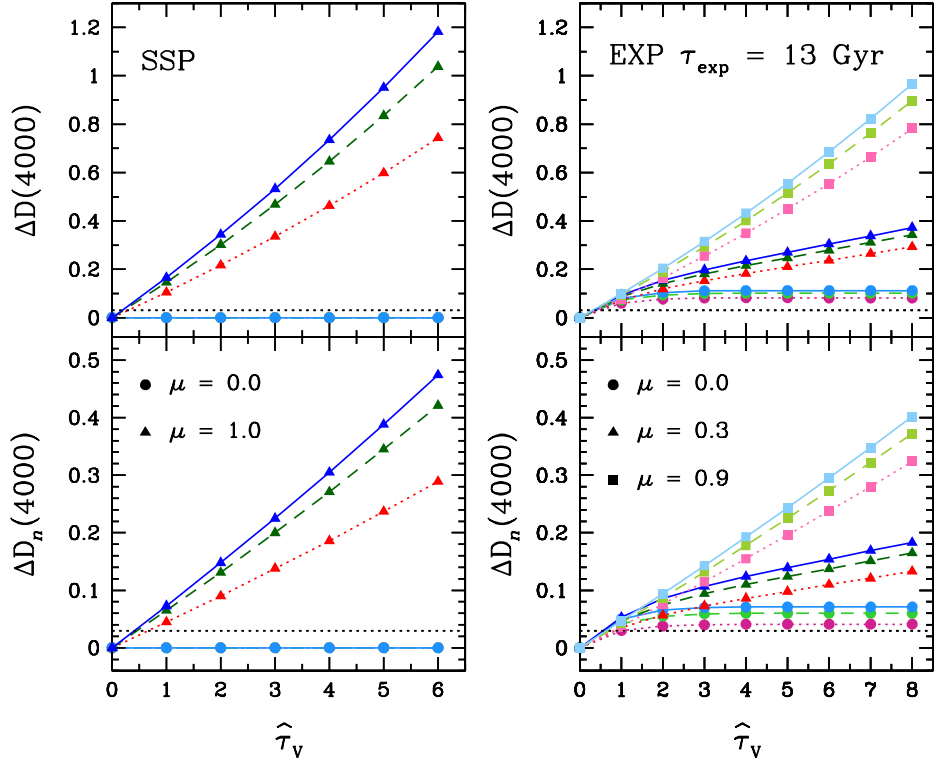} 
\vspace{- 0.7in}
\caption{D(4000) [top panels] and \dnf\ [bottom panels] differences, 
         $\Delta{index} \equiv 
         index$(\taueff$_{V}) - index($\taueff$_{V} = 0)$, as a 
         function of \taueff$_V$ for SSPs [left panels] and for an exponential 
         SFH with $\tau_{exp} = 13$ Gyr [right panels].  Different 
         ages are represented by: 0.5 (dotted lines; red shades), 
         5 (dashed lines; green shades),
         \& 13 Gyr (solid lines; blue shades).  The black horizontal 
         dotted lines represent the typical measurement error.
         \label{fig:d4000}}
\end{center}
\end{figure*}

Figure~\ref{fig:d4000} shows the response of D(4000) [top panels] and 
\dnf\ [bottom panels] to attenuation from dust for SSPs [left panels] and
exponential SFHs with $\tau_{exp}=13$~Gyr [right panels].  The narrower 
index is certainly less affected by dust (note the different y-axis scales),
but it is by no means impervious to dust reddening effects, and can reach
deviations with magnitudes much larger than the typical error for this index.
Dust effects are smaller at younger ages, but they can be significant at 
all ages for realistic values of \taueff$_{V}$.  A detailed discussion of
the influence of these dust effects on the determination of ages and 
metallicities is deferred to appendix \ref{sec:fits}.

Gorgas \etal\ (1999) suggest that D(4000) could be corrected for the
effects of internal reddening according to,
\begin{equation}
{\mbox{D(4000)}}_{\mbox{\scriptsize corrected}} = 
{\mbox{D(4000)}}_{\mbox{\scriptsize observed}}*10^{-0.0988E(B-V)},
\label{eq:gorgas}
\end{equation}
derived using the mean extinction curve from Savage \& Mathis (1979).
However, this assumes that the dust is distributed as a screen, which
is likely unrealistic for galaxies (\eg\ Witt, Thronson, \& Capuano 1992),
and that the value of the color excess $E(B-V)$ is known.
Even knowing $E(B-V)$, as we do for these models, the above formula does not 
reproduce the dust-free D(4000) value for the CF00 dust model.  The 
numerical constant in equation~(\ref{eq:gorgas}) could be adjusted to 
obtain a good 
fit, but this number is a function of age, depends on the dust properties,
and requires {\it a priori} knowledge of $E(B-V)$.  Hence, we do not explore
this putative correction
further, but note that, as we concluded for the Lick indices, there is 
likely no such simple dust correction for the 4000~\AA\ break of mixed stellar 
populations and complex dust geometries.

\subsection{Dust Sensitivity of the \ion{Ca}{2} Triplet Indices}\label{sec:CaT}

\begin{figure}
\begin{center}
\includegraphics[bb=110 144 512 718, width=3.5in]{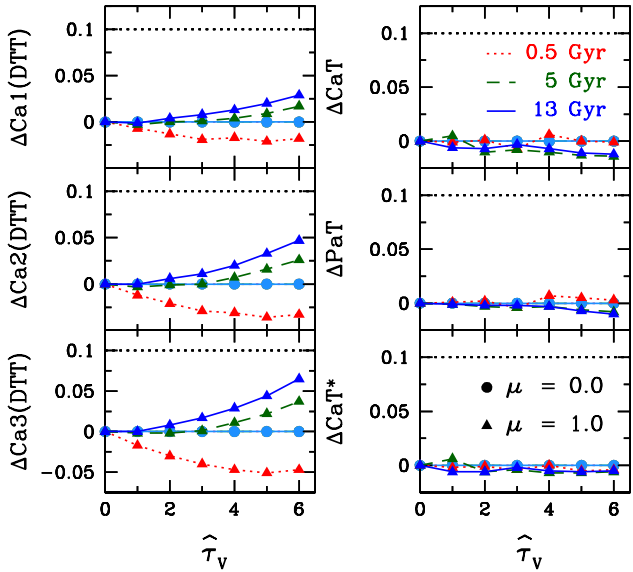} 
\vspace{- 1.95in}
\caption{Calcium triplet index differences, $\Delta{index} \equiv 
         index$(\taueff$_{V}) - index($\taueff$_{V} = 0)$, as a function of 
         \taueff$_V$ for solar metallicity SSPs.  Different 
         ages are represented by: 0.5 (dotted lines; red shades), 
         5 (dashed lines; green shades), 
         \& 13 Gyr (solid lines; blue shades).
         Different values of $\mu$ are represented as 0.0 (circles), 
         and 1.0 (triangles). The black horizontal dotted lines 
         represent typical measurement errors for the different indices.
         \label{fig:CaT_SSPs}}
\end{center}
\end{figure}

\begin{figure}
\begin{center}
\includegraphics[bb=110 144 512 718, width=3.5in]{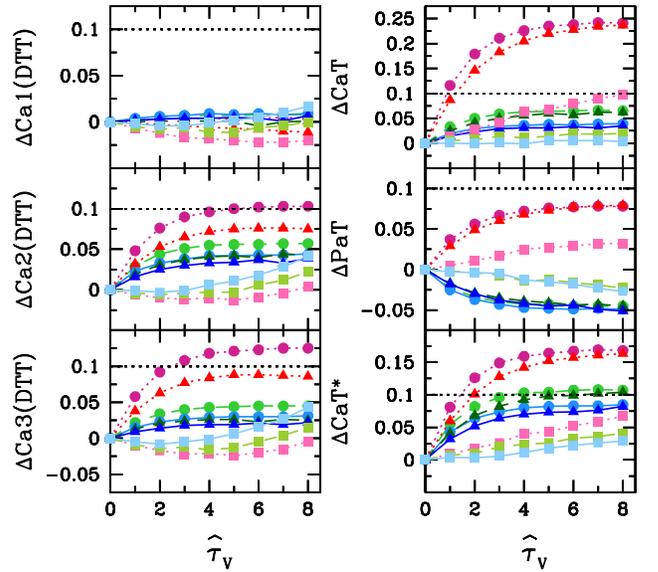} 
\vspace{- 1.95in}
\caption{Same as Figure~\ref{fig:CaT_SSPs} but for for an exponential SFH 
         with $\tau_{exp} = 13$ Gyr.  Different
         values of $\mu$ are represented by 0.0 (circles), 0.3 (triangles),
         and 0.9 (squares).
\label{fig:CaT_expsfh}}
\vspace{0.1in}
\end{center}
\end{figure}

The response of the \ion{Ca}{2} triplet indices of DTT and 
Cenarro \etal\ (2001a) as a function of \taueff$_{V}$, $\mu$, 
and age for solar metallicity SSP models are shown in Figure~\ref{fig:CaT_SSPs}
and for an exponential SFH with $\tau_{exp}=13$~Gyr in 
Figure~\ref{fig:CaT_expsfh}.
The three classical Lick-style Ca indices of DTT
(left) show some sensitivity to dust extinction for the SSPs, Ca3 showing the 
largest sensitivity, but the deviations remain below
the typical measurement errors thus would not be detected.  The generic CaT 
indices (right) are virtually unaffected by dust extinction in the SSP models
at any age.  As usual, the situation is not as straightforward for the 
exponential SFH (Fig.~\ref{fig:CaT_expsfh}).  
Here, the only index that is 
not significantly affected is the Ca1 index of DTT.  All of the generic 
indices of Cenarro \etal\ (2001a) are significantly
altered due to dust extinction, particularly at young ages.  
As in the case for the Lick indices in the exponential SFH models 
(\S\ref{sec:lick}, Fig.~\ref{fig:expsfh}), dust effects are often stronger 
for models with smaller $\mu$ and the effects saturate at optical depths 
above about \taueff$_{V}=4$.  In general, though, for models that do not have 
significant amounts of current SF, the dust reddening effects in the 
\ion{Ca}{2} triplet indices are still small compared to measurement errors,
and should not cause problems in their interpretation.

\subsection{Dust Sensitivity of the Rose Indices}\label{sec:rose}

\begin{figure*}
\begin{center}
\includegraphics[width=5.0in]{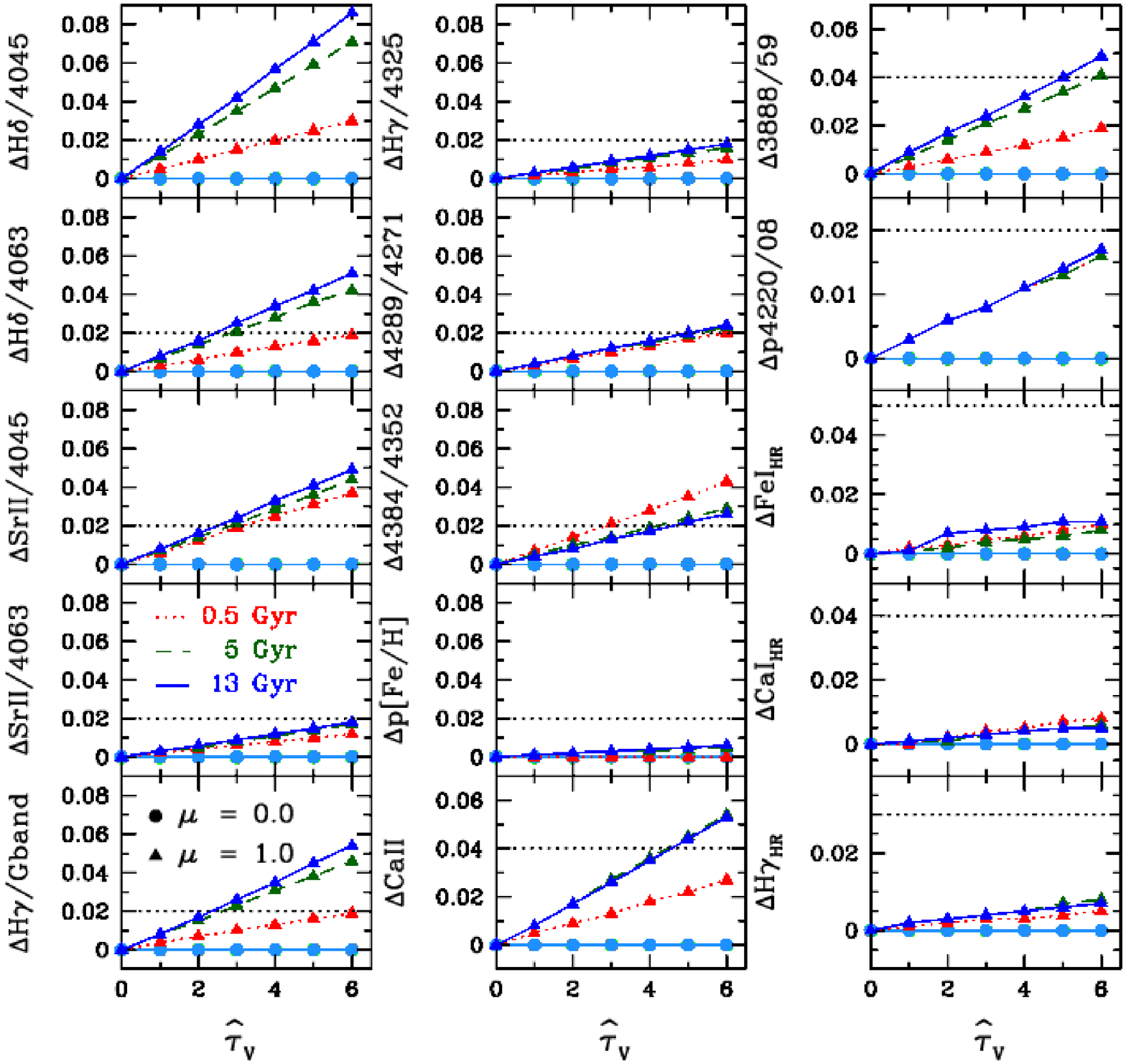} 
\vspace{- 0.1in}
\caption{Rose index differences, $\Delta{index} \equiv 
         index$(\taueff$_{V}) - index($\taueff$_{V} = 0)$, as a function 
         of \taueff$_V$ for solar metallicity SSPs.  Different 
         ages are represented by: 0.5 (dotted lines; red shades), 
         5 (dashed lines; green shades),
         \& 13 Gyr (solid lines; blue shades).
         Different values of $\mu$ are represented by 0.0 (circles), 
         and 1.0 (triangles). The black horizontal dotted lines 
         represent typical measurement errors for the different indices.
         \label{fig:Rose_SSPs}}
\end{center}
\end{figure*}

\begin{figure*}
\begin{center}
\includegraphics[width=5.0in]{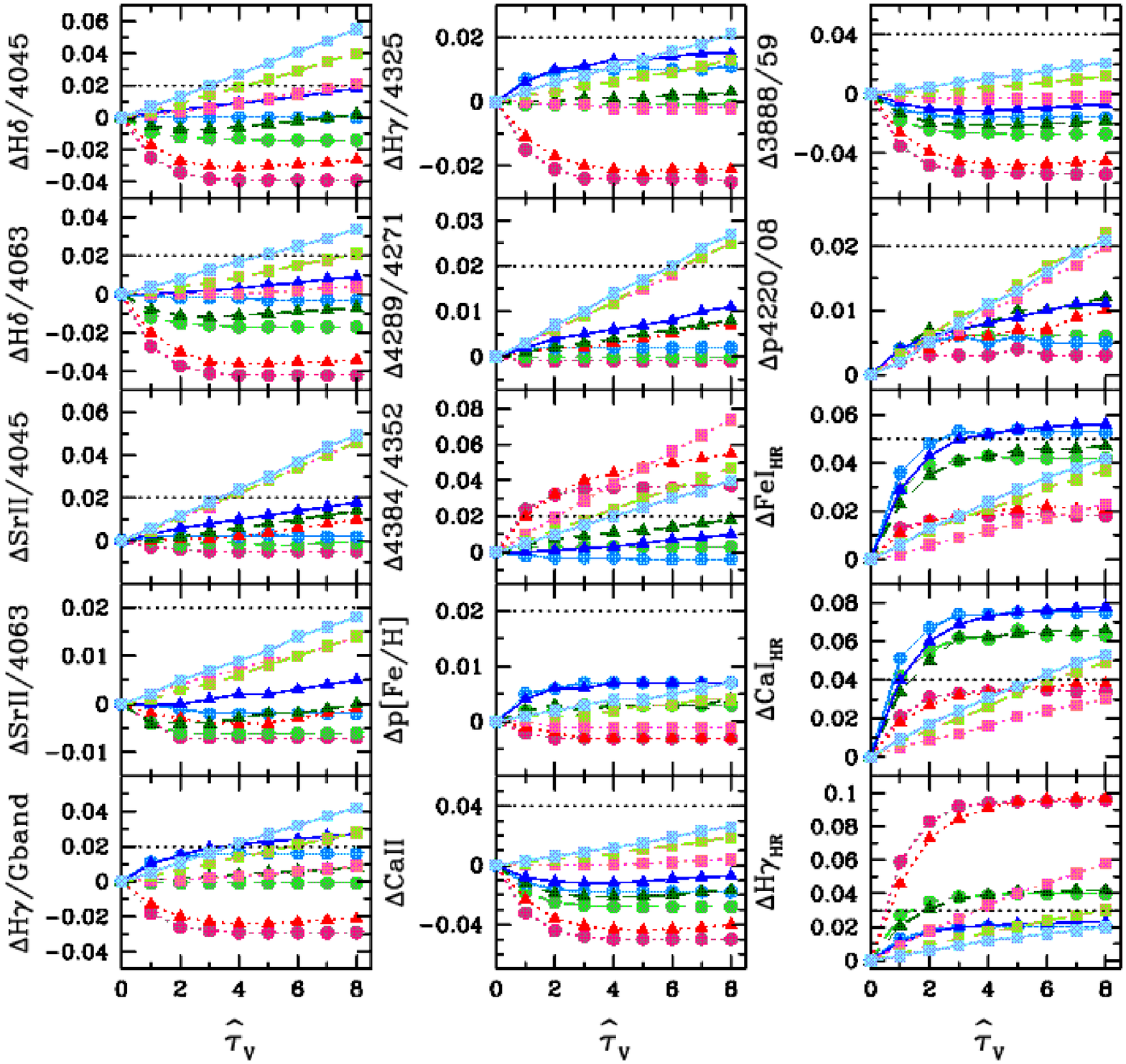} 
\vspace{- 0.1in}
\caption{Same as Figure~\ref{fig:Rose_SSPs} but for
         an exponential SFH with $\tau_{exp} = 13$ Gyr. 
         Different
         values of $\mu$ are represented by 0.0 (circles), 0.3 (triangles),
         and 0.9 (squares).\label{fig:Rose_expsfh}}
\end{center}
\end{figure*}

The response of the Rose (1984, 1994) and Jones \& Worthey (1995) indices
are shown in Figure~\ref{fig:Rose_SSPs} for SSPs and in 
Figure~\ref{fig:Rose_expsfh} for an exponential SFH with $\tau_{exp}=13$~Gyr.
Symbols and line types are as in Figures~\ref{fig:SSPs} \& \ref{fig:expsfh}.
For SSP models there are a few indices that are considerably modified by dust 
reddening, namely the H$\delta$/\ion{Fe}{1}$\lambda$4045, 
H$\delta$/\ion{Fe}{1}$\lambda$4063, 
\ion{Sr}{2}/\ion{Fe}{1}$\lambda$4045, H$\gamma$/Gband, and \ion{Ca}{2} indices,
particularly at older ages.  All other indices, including the 
pseudo-equivalent width indices, are essentially unaffected by the dust
reddening in the SSPs at any age.  For the exponential SFH, shown in 
Figure~\ref{fig:Rose_expsfh}, the response of the indices gets quite 
complicated.  The offset from the dust-free case can be either positive 
(typically for the older models) or negative (typically for youngest models). 
Again, for the $\mu=0.0$ models and for some of the $\mu=0.3$ models, the
index offsets saturate for optical depths above \taueff$_{V}=3$.
Ultimately though, even with current SF, for the majority of the Rose indices,
the index offsets due to dust reddening are modest.
On the other hand, with a significant amount of current SF, the 
pseudo-equivalent width indices can be quite affected, more so at young ages 
for H$\gamma_{{\mbox{\scriptsize HR}}}$, but more so for old ages for 
\ion{Ca}{1}$_{{\mbox{\scriptsize HR}}}$ and 
\ion{Fe}{1}$_{{\mbox{\scriptsize HR}}}$.  The ``HR'' 
indices are virtually unaffected for SSP models at all ages.

\section{Discussion} \label{sec:disc}

\begin{figure}
\begin{center}
\includegraphics[bb=48 144 592 718,width=3.3in]{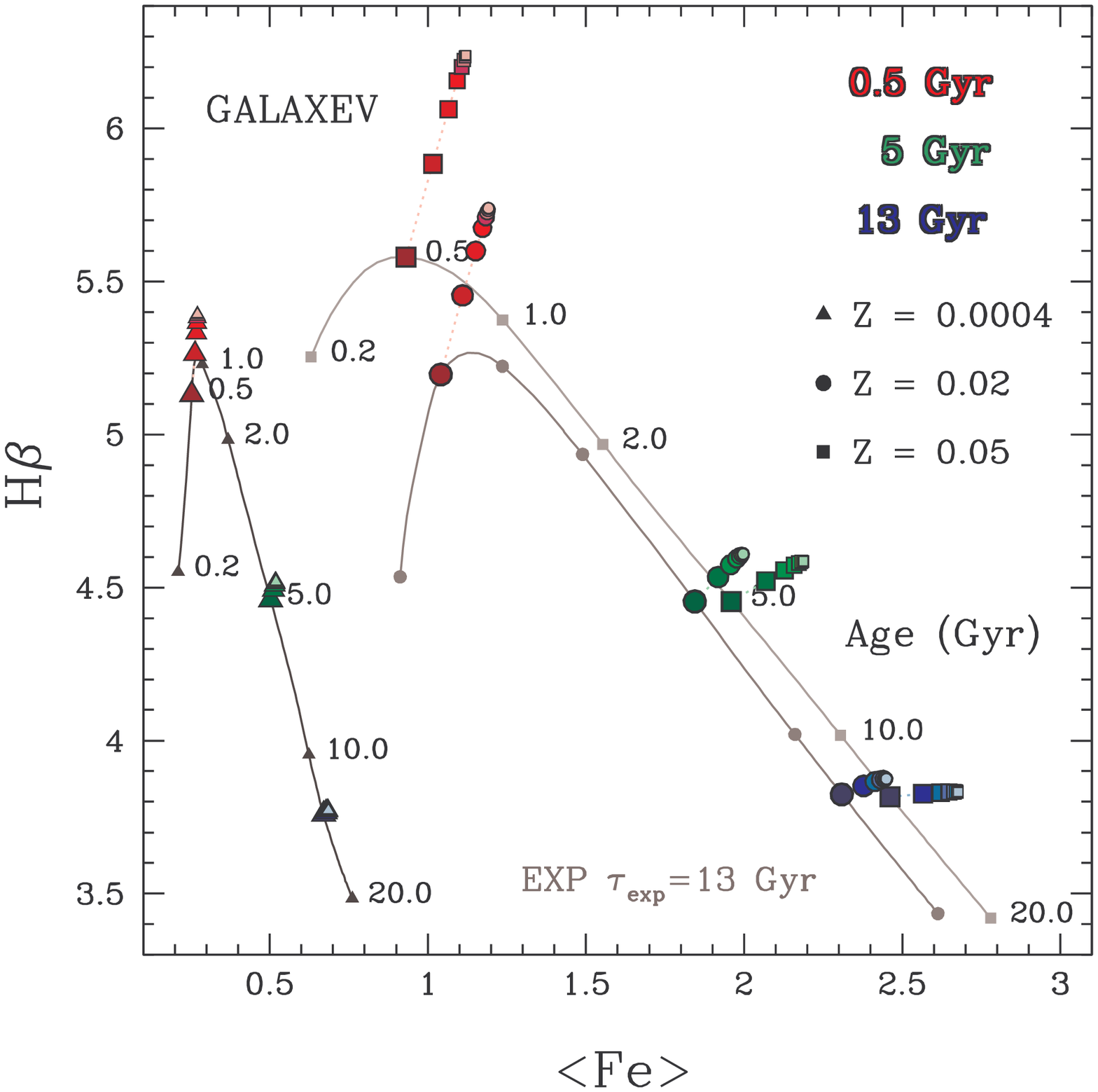} 
\caption{H$\beta$ versus \avgFe\ for dust-free exponential SFH models with 
        $\tau_{exp} = 13$ Gyr and $\mu = 0.3$ for three metallicities (gray
        curves), Z = 0.0004 (triangles), 0.02 (circles), and 0.05 (squares).  
        Ages are labeled along
        the model tracks at 0.2, 0.5, 1.0, 2.0, 5.0, 10.0, and 20.0 Gyr.
        The colored points (red, green, and blue shades for 0.5, 5, and 13 Gyr
        respectively) are the model indices measured
        for the same exponential SFH models, but with dust extinctions from
        \taueff$_V = 0$ (largest point size) to \taueff$_V = 8$ 
        (smallest point size).
         \label{fig:Hb_Fe}}
\end{center}
\end{figure}

As an illustration of the potential dangers of dust extinction when using 
absorption-line indices to determine ages and metallicities of star forming
galaxies, in Figure~\ref{fig:Hb_Fe} we present model tracks for the 
Lick H$\beta$ versus \avgFe\ diagnostic plot for the dust-free exponential 
SFH with $\tau_{exp} = 13$~Gyr for $Z=0.0004$, 0.02, and 0.05 (gray curves).  
The circles are the index values for the same SFH, but with dust extinctions 
ranging from \taueff$_{V} = 0$ (largest point size) to \taueff$_{V} =8$ 
(smallest point size).  There is virtually no error for old and metal-poor 
galaxies, but significant problems can arise at higher metallicities.  For 
example, the location in the H$\beta$ versus \avgFe\ plane for a solar 
metallicity stellar population at any age with \taueff$_{V} = 1$ lies closer 
to the $Z=0.05$ model curve.  The
age errors are not significant in this case, but as extinction
increases, the data points lie significantly outside the region covered 
by the model grids.  

However, since in reality we don't know the true SFH of a galaxy, what is
often computed are ``SSP-ages and metallicities'', \ie, the values that
would be measured if a given index--index combination is fit to SSP model 
grids.  In order to gauge the magnitude of the errors
on SSP ages and metallicities due to dust extinction, we fit the model 
galaxies to SSP grids of different index--index combinations.  The error on
the derived parameters is taken as the difference between the 
fit for the dusty model and the corresponding dust-free model (\ie\ 
$\Delta$A = Age(\taueff$_{V}$) $-$ Age(\taueff$_{V}$ = 0)).  In some cases
we encounter difficulties with the models with current SF, as these models
can lie off the SSP model grids, even in the dust-free models.  Additionally,
due to the double-valued nature of the Balmer lines at young ages (discussed 
in \S\ref{sec:lick}), there is an additional degeneracy for points that lie
in the youngest region of the SSP model grids.  This was the case for all of 
the exponential SFH models at 0.5 Gyr, for which a large
fraction of the stars are very young.  Details of the fits and tables providing
the resulting age and metallicity errors for a selected number of index--index 
diagnostic plots are provided in Appendix~\ref{sec:fits}.  The primary result
from this analysis is that, in general, when dust extinction affects the
SSP age and metallicity determinations, the errors on the physical parameters
are of the same order as the 1$\sigma$ confidence intervals from typical
measurement errors, and thus would not likely be detected above the noise.  
The most notable exception is any index--index plane using the \dnf, which
suffers significant errors even in the SSP models.  Index--index grids using
the longest baseline indices can also suffer significant errors on the 
measured physical parameters due to the effects of dust.

\begin{figure}
\begin{center}
\includegraphics[bb=48 144 592 718,width=3.3in]{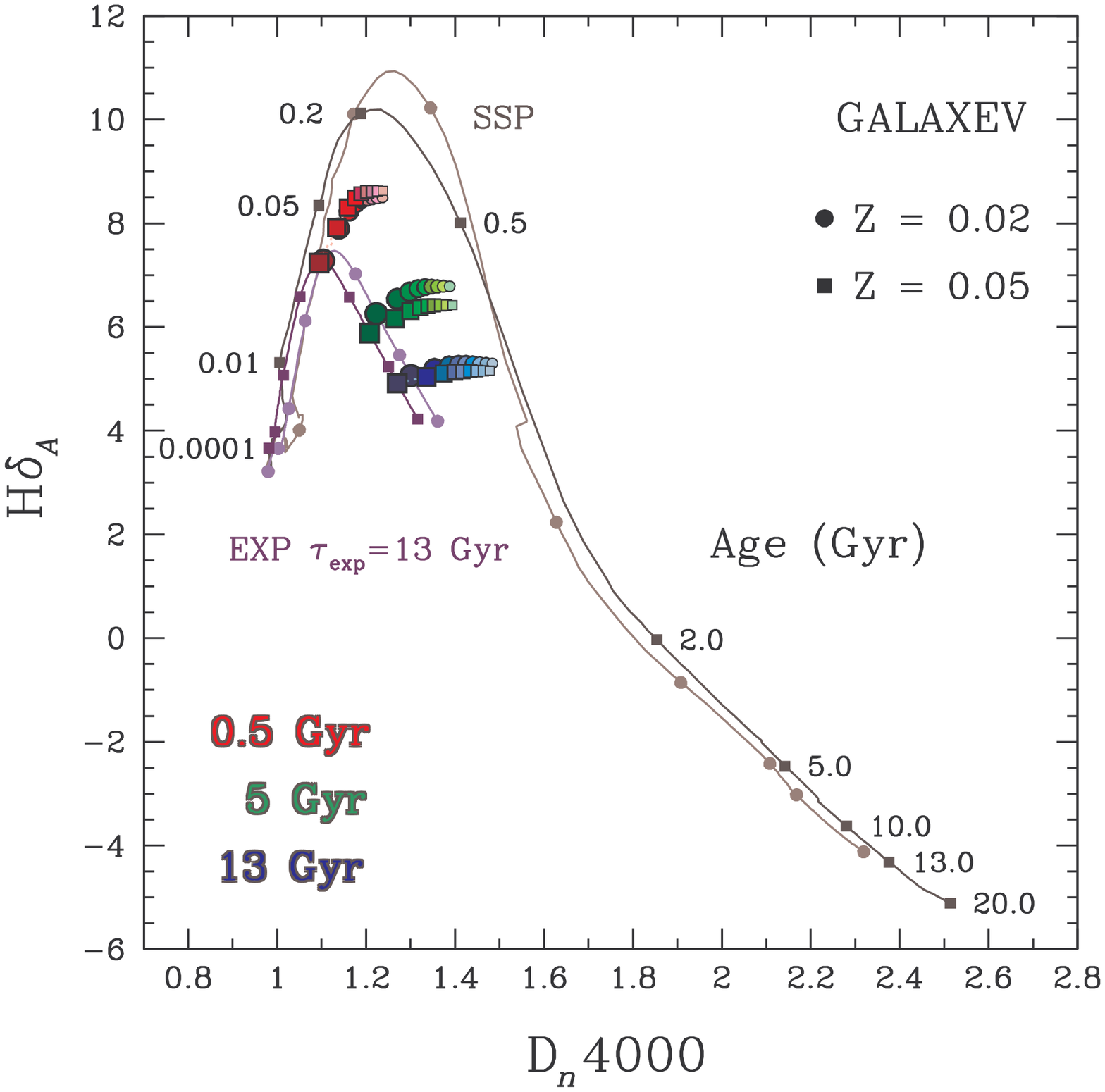} 
\caption{H$\delta_{A}$ versus \dnf\ for dust-free SSPs of solar (circles)
         and Z=0.05 (squares)
         metallicity (gray curves) and an exponential 
         SFH with $\tau_{exp} = 13$ Gyr for the same two metallicities
         (purple curves).  Ages are labeled along the model tracks at
         0.0001, 0.01, 0.05, 0.2, 0.5, 1.0, 2.0, 5.0, 10.0, 13.0, and 20.0 Gyr.
         The colored points (red, green, and blue shades for 0.5, 5, and 13 Gyr
         respectively) are the model indices measured
         for exponential SFH models of the same two metallicities, but with 
         dust extinctions from \taueff$_V = 0$ (largest point size) to 
         \taueff$_V = 8$ (smallest point size).
         \label{fig:HdelA_d4000}}
\end{center}
\end{figure}

Figure~\ref{fig:HdelA_d4000} illustrates the potential pitfalls of 
discounting dust extinction in an analysis that uses index diagnostics to 
determine SFHs.
There, we plot H$\delta_{A}$ versus \dnf\ for the dust-free 
SSP (gray curves) and 
exponential SFH (purple curves) models for solar and $Z=0.05$ metallicity
and ages up to 20~Gyr.  The triangles represent the $\mu=0.3$ model indices 
for \taueff$_V = 0$ (largest point size and darkest shade) to \taueff$_V = 8$ 
(smallest point size, lightest shade).  The locus of galaxies in the 
H$\delta_{A}$ -- \dnf\ plane has been used to determine their SFHs
(\eg\ Kauffmann \etal\ 2003a).  Clearly, if a galaxy 
contains a significant amount of dust, its location in this plane will be
best fit by a SFH that is different from its true SFH.  As shown,
the exponential SFHs move closer to the SSP region when dust is added.

Fortunately, in the Rose dwarf/giant diagnostic plot
(\ion{Sr}{2} $\lambda$4077/\ion{Fe}{1}$\lambda$4063
versus H$\delta$/$\lambda$4063), the dust effects run
along the lines of constant surface gravity.  Thus, this diagnostic
plot is still useful for determening the relative amounts of young versus 
evolved stars, even when dust is present.

It must be emphasized that our analysis has not accounted for emission-line 
contamination.  Dust effects aside, this could potentially be 
the most worrisome pollutant for all age determinations involving Balmer lines,
particularly for the youngest stellar populations.
While considering the higher-order Balmer lines (H$\gamma$ and H$\delta$) 
greatly reduces the problem, it does not go away altogether.  

To summarize, we have considered dust effects on absorption-line indices using 
the Bruzual \& Charlot (2003) population synthesis models incorporating the 
multi-component model of CF00 for the line and continuum attenuation due to 
dust.  For quiescent stellar populations (\eg\ spheroids and globular 
clusters), dust extinction effects are
small for most indices.  A notable exception is the 4000~\AA\ break,
whose sensitivity to dust can translate into significant errors in the age
determination of the stellar population.  For models with current SF, many of 
the indices are significantly
modified due to dust reddening effects, and their behavior depends on age, 
dust distribution, and the effective optical depth.  Unfortunately, no 
simple dust-correction can be prescribed, but future spectroscopic studies of
stellar populations in dusty environments (\eg\ late-type spirals) ought to 
consider possible effects due to dust attenuation in their measurements and 
its resulting effect on physical interpretations.  Fortunately, there are
particular indices that are negligibly affected by dust extinction (compared
to the current level of measurement precision) at different regions in the
physical parameter space, and the safest current approach consists of mapping 
a broad range of indices with the shortest wavelength baselines.

\acknowledgments

I am most grateful to St{\' e}phane Courteau for his continued advice and 
support.  Special thanks go out to Jim Rose, Ricardo Schiavon, Scott Trager, 
and Guy Worthey for their careful readings of an early version 
of the manuscript, and to Jes{\' u}s Gonz{\' a}lez for useful discussions.
I would also like to thank the anonymous referee for a detailed report
that helped improve this paper.
The author acknowledges financial support from the National Science and 
Engineering Council of Canada.


\begin{appendix}

\section{Age and Metallicity Fitting} \label{sec:fits}

In order to translate absorption-line indices to a physical age and 
metallicity scale, the index measurements must 
be compared with stellar population synthesis models.  In their most basic 
form, commonly referred to as simple stellar populations (SSPs), these models 
provide evolutionary information for a coeval population of stars born with 
a given composition and initial mass function (IMF).  Several such SSP models 
have been produced by a number of independent groups and are in a constant 
state of flux as improvements to many of the input parameters 
(\eg\ stellar libraries, model atmospheres, convection, mass loss, mixing) 
come to light.  There are discrepancies among the different models that, 
depending on the application, may result in significantly different 
interpretations of the observed stellar population signatures.  We are
not interested here in a detailed comparison of the different SPS models, 
but rather seek a quantitative guide for the magnitude of the errors 
due to dust effects on the physical parameters derived from absorption-line 
indices.  To this end, we use the 2003 implementation of the Bruzual \& 
Charlot (2003) SPS models (GALAEX) for the age and metallicity determinations, 
the same as are used to compute the model ``galaxies'' 
(see \S\ref{sec:models}).

While most galaxies have likely undergone SFHs that are much more complex 
than the simple SSP described above, the common practice for line-index 
measurements is to compute ``SSP-ages and metallicities'', that is, compare 
the galaxy indices to the SSP grids, irrespective of the true SFH of the 
galaxy (which is unknown).  We adopt this technique here and thus refer to 
SSP parameters even for the models with exponential SFHs.

Ages and metallicities are determined by fitting the model galaxy index 
measurements to the SSP model grids using a maximum-likelihood approach.
We retain the same grid in time as is provided in the model SSPs (220 
unequally spaced time steps from $t$ = 0--20 Gyr)
and interpolate (linearly) between the model metallicities on a fine grid
of 120 unequally spaced steps in $Z$.  
In order to compare all indices on a similar scale, the atomic indices 
computed as equivalent widths in \AA, $I_{EW}$,
are converted to a magnitude, $I_{mag}$, as:

\begin{equation}
I_{mag} = -2.5\,\log_{10}\left(1-\frac{I_{EW}}{\Delta\lambda}\right)
\end{equation}
 
where $\Delta\lambda$ is the width in \AA\ of the index feature bandpass.  
The corresponding error in magnitude is:
\begin{equation}
\sigma_{mag} = \frac{2.5\,\log_{10}(e)}{\Delta\lambda\,10^{-0.4I_{mag}}}\,\sigma_{EW}.
\end{equation}

The two definitions of the 4000 \AA\ break, which are dimensionless flux 
ratios, are converted to a magnitude scale as:
\begin{equation}
\mbox{D4000}_{mag} = -2.5\,\log_{10}\left({\mbox{D4000}_{ratio}}\right)
\end{equation}
 
with the corresponding error in magnitude as:
\begin{equation}
\sigma[\mbox{D4000}_{mag}] = 2.5\,\log_{10}(e)\frac{\sigma[\mbox{D4000}_{ratio}]}{\mbox{D4000}_{ratio}}.
\end{equation}

We compute an age and metallicity by minimizing the following figure of merit:

\begin{equation}
FOM = \frac{1}{N-1}\sum_{i=1}^{N}\frac{[O_{i}-M(A,Z)_{i}]^2}{\delta O_{i}^{2}},
\end{equation}

where $N$ is the number of indices (only 2 are used here, but this method 
could be generalized to fit multiple indices simultaneously), $O_i$ is the 
``observed'' value (in magnitudes) of index $i$ and $\delta O_{i}$ is its 
error in magnitudes, and $M(A,Z)_{i}$ is the SSP model value of index $i$ for 
a given age and metallicity combination.

\begin{figure}
\vspace{0.25in}
\plottwo{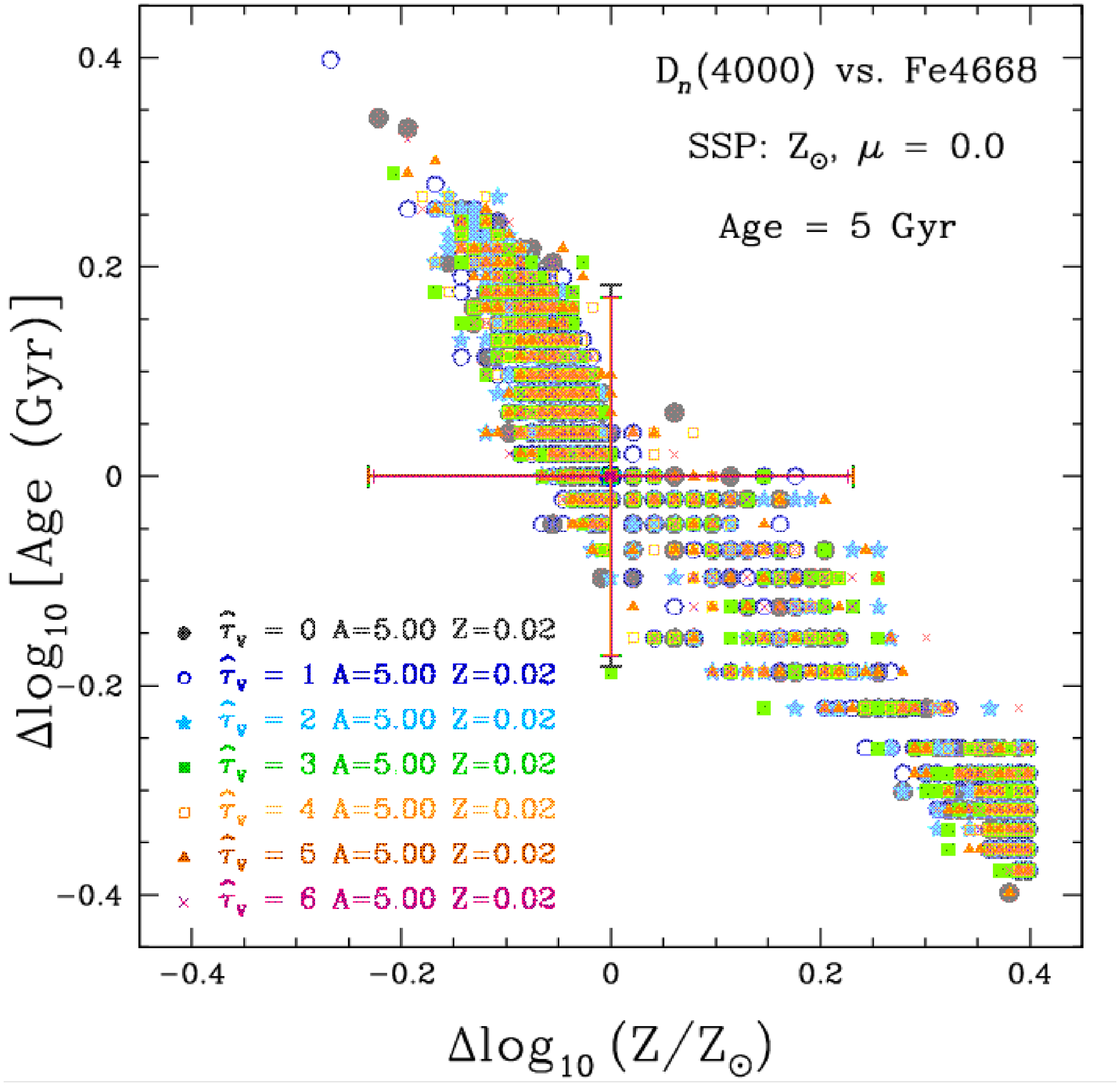}{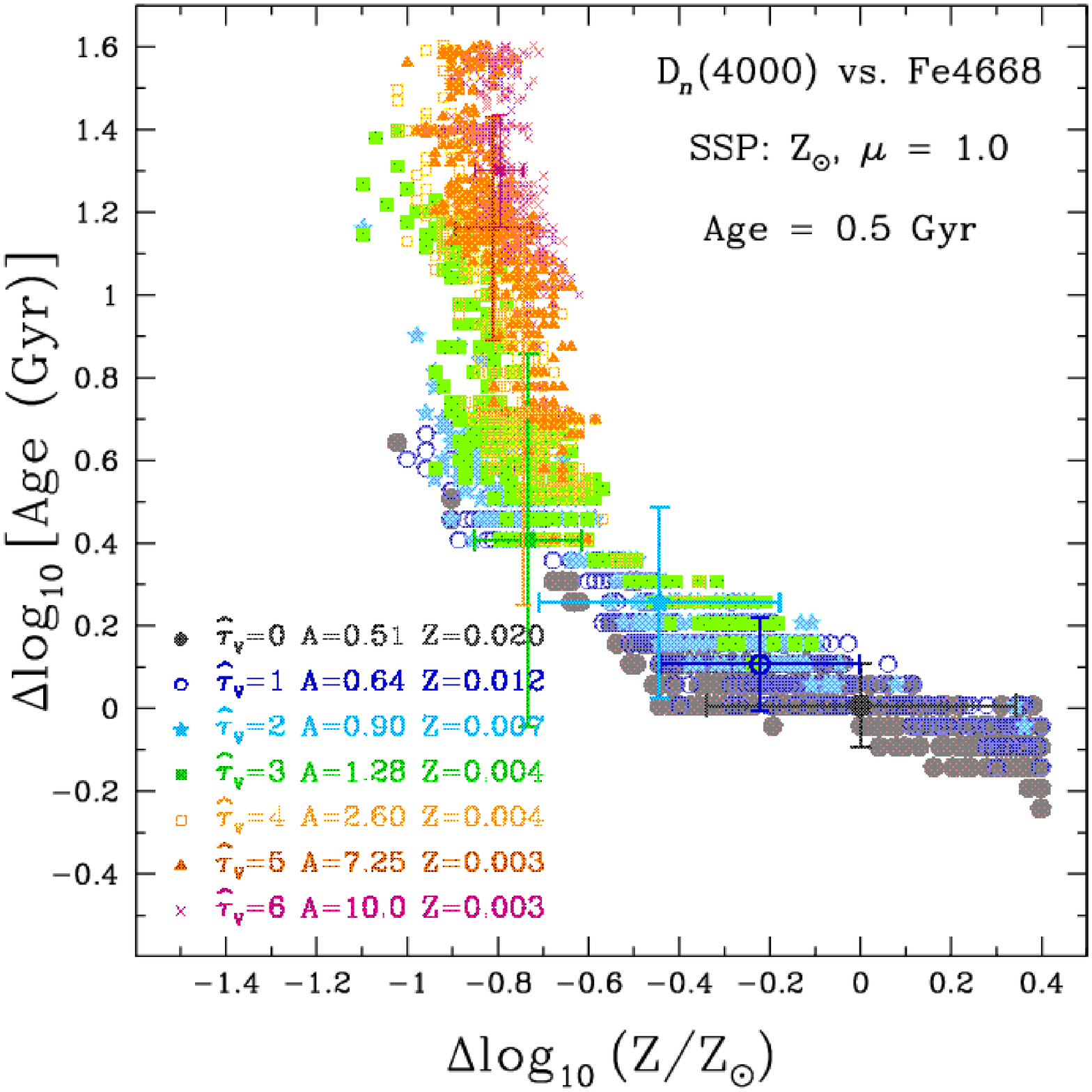} 
\caption{Examples of the Monte Carlo fits for \dnf\ versus Fe4668 diagnostic 
         plots, represented as the logarithmic
         difference between the physical parameter derived from each
         simulated point relative to the best-fit value (also see the 
         corresponding index--index plot in Figure~\ref{fig:indexgrids}, 
         right panel). The error bars represent the 1$\sigma$ confidence 
         limits.  Left: SSP model at 5 Gyr
         with $\mu = 0.0$. Right: SSP model at 0.5 Gyr with $\mu = 1.0$ 
         (note the different y-axis scales).
         \label{fig:montecarlo}}
\end{figure}

\begin{figure}
\vspace{-0.25in}
\plottwo{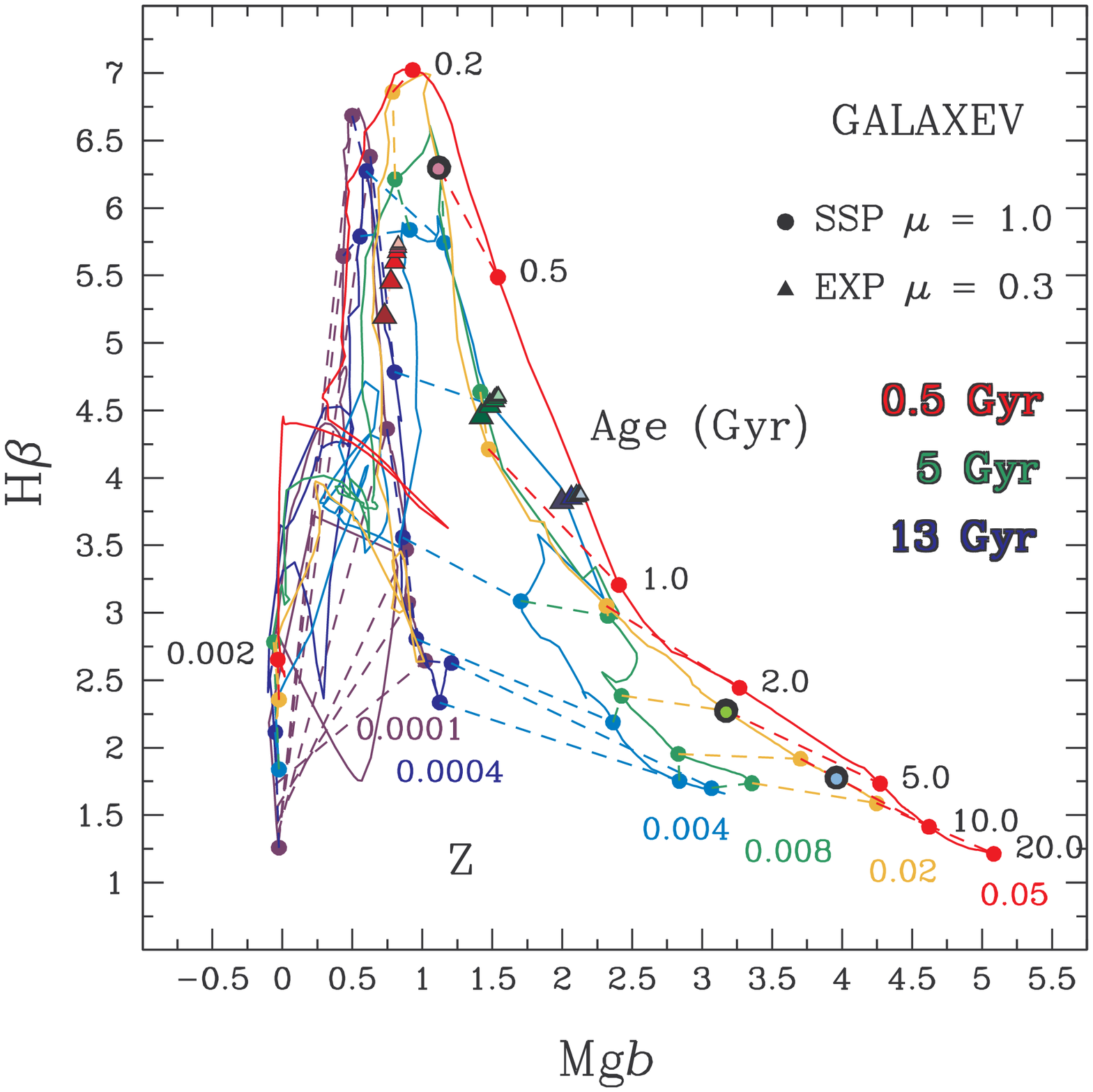}{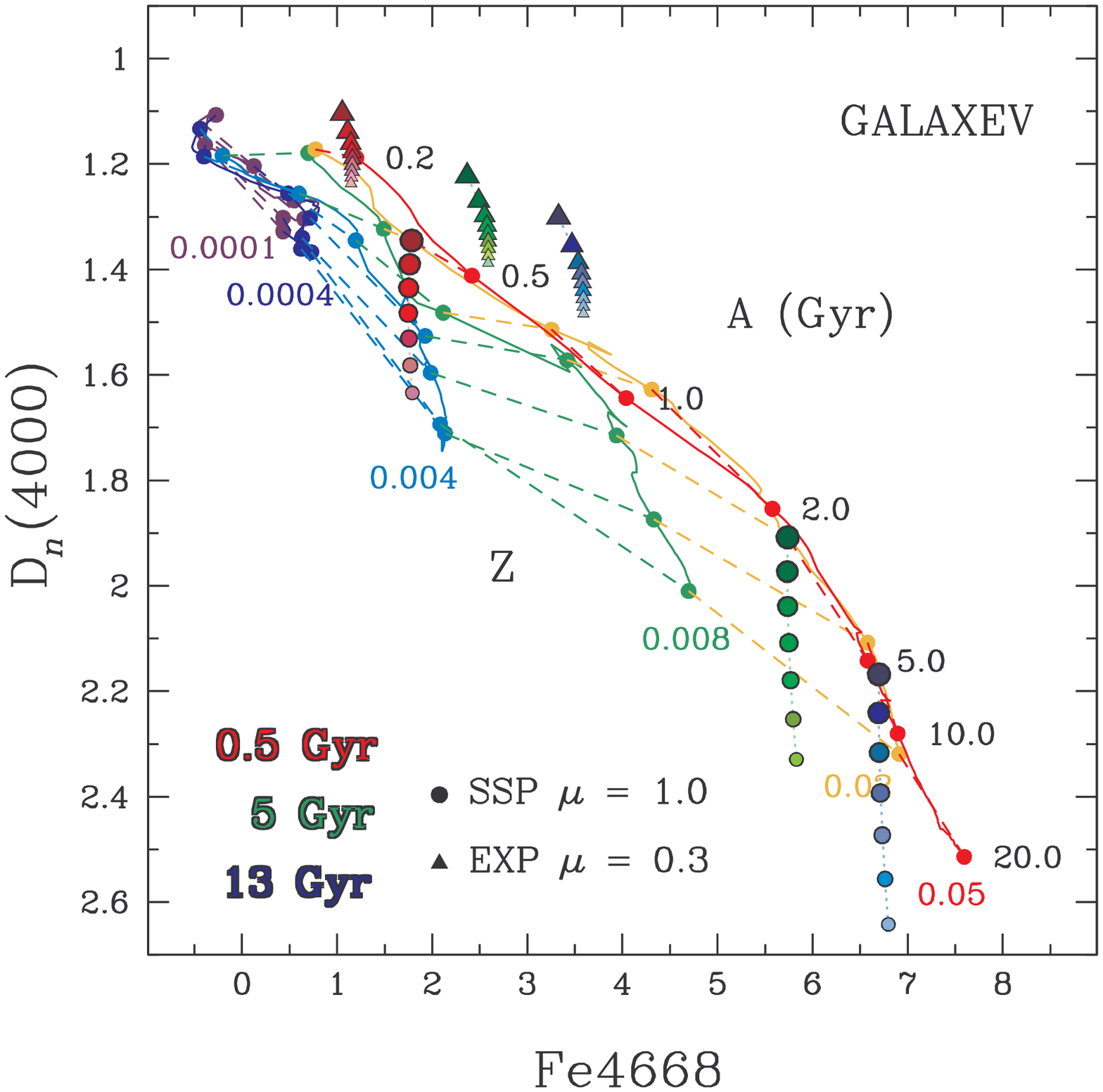} 
\caption{H$\beta$ versus Mg{\it b} [left] and \dnf\ versus Fe4668 [right]  
        diagnostic plots.  The grids are dust-free BC03 
        SSP models.  Lines of constant age (dashed) are shown for ages 
        0.2, 0.5, 1.0, 2.0, 5.0, 10.0, and 20.0 Gyr in both panels, but in
        the H$\beta$ versus Mg{\it b} plot (left), the age extends to 
        0.002 Gyr to illustrate the messy young age, double-valued Balmer
        line, region (see text).  Lines of constant
        metallicity (solid) are shown for Z = 0.0001 (purple), 
        0.0004 (dark blue), 0.004 (light blue), 0.008 (green), 0.02 (yellow),
        and 0.05 (red).  The circles are the indices measured from the 
        solar metallicity SSP models with
        $\mu = 1.0$ for \taueff$_V$ = 0 (largest point size) to 
        \taueff$_V = 6$ (smallest point size),  and the triangles 
        are the exponential SFH model indices with $\tau_{exp} = 13$ Gyr and
        $\mu = 0.3$ for \taueff$_V = 0$ (largest point size) to 
        \taueff$_V = 8$ (smallest point size).  The red, green, and blue
        shades are for model ages of 0.5, 5, and 13 Gyr respectively.
         \label{fig:indexgrids}}
\end{figure}

Errors for the individual age and metallicity measurements are estimated using 
a Monte Carlo approach.  One thousand realizations of the model fits are 
computed using errors drawn from a normal distribution of the observational 
errors (taken here as the ``typical'' observation errors for each index 
quoted in \S\ref{sec:lick}).  The errors for the measured ages and 
metallicities are taken as half the interval containing 68\% 
of the 1000 Monte Carlo realizations (\ie\ the 1$\sigma$ confidence interval).
Two examples of the fits are given in Figure~\ref{fig:montecarlo} which 
shows the logarithmic difference between the physical parameters derived 
from each simulated point relative to the best-fit value for the 
\dnf\ -- Fe4668 plane (see also 
the corresponding index--index plot in Figure~\ref{fig:indexgrids} 
[right panel]).    The left panel presents the fit for the dust-free 
($\mu = 0.0$) SSP case which indeed finds the 
correct model age (A = 5 Gyr; $\Delta \log_{10}$[Age (Gyr)] = 0) and 
metallicity (Z = 0.02; $\Delta[\log_{10}(Z/Z_\odot)] = 0$), with corresponding 
1$\sigma$ confidence intervals (represented as error bars)
of $\sim$ 0.2 dex for both parameters.
The right panel presents the fit for the dusty case ($\mu = 1.0$) at 0.5 Gyr
(but note the different y-axis scales).  The \taueff$_{V}= 0$
fit finds the correct physical parameters, but as the dust opacity increases,
the best fit wanders around in age 
(from A = 0.5 Gyr at \taueff$_{V}= 0$ to A = 10.0 Gyr at \taueff$_{V}= 6$)
and metallicity (from $Z$ = 0.02 at \taueff$_{V}= 0$ to $Z$ = 0.003 at 
\taueff$_{V}= 6$).  Since we do not consider
extrapolations, there are ``edges'' in these plots that represent the limits
of the model grids.  As a result, the confidence limits can be underestimated
for points fit near the model grid limits, which also generally correspond to 
a region of greater degeneracy in age and metallicity.  As a result, any 
points that lie near the model limits should be treated with caution.

Given that we are comparing model galaxies with exponential SFHs against 
SSP grids, it is inevitable that some of the parameter space will fall 
outside the model grids and thus cannot be fit reliably.  In fact, even for 
some of the SSP models, the dust extinction can be severe enough to push the 
locus of the model point in the index--index diagram off the SSP model grids.  
Both of these situations are observed in the \dnf\ versus Fe4668 index--index
diagram in Figure~\ref{fig:indexgrids} [right panel].

Tables~\ref{tab:errs_SSP} \& \ref{tab:errs_EXP} present the results from our 
age and metallicity fitting for a selected number of index--index combinations:
H$\beta$ versus Mg{\it b}, Mg$_2$, and [MgFe]$\arcmin$ (top) and 
H$\beta$, H$\delta_{A}$, and \dnf\ versus Fe4668 (bottom) for the SSP and 
exponential SFH model galaxies respectively (tables for other index 
combinations are available from the author upon request).  For each 
index-pair, we list the model age, {\small MA}, the effective $V$-band optical 
depth, \taueff$_{V}$, whether the fit is reliable in the ``fit'' column 
(see below), the best-fit age with its 1$\sigma$ error in brackets, the 
difference between the measured age in the extincted models and the dust-free 
age, $\Delta$A $\equiv$ Age(\taueff$_{V}$) $-$ Age(\taueff$_{V}$ = 0), and 
the same for the metallicities,  which are given as $\log_{10}(Z/Z_\odot)$ and
$\Delta Z \equiv$ $\log_{10}(Z/Z_\odot)$(\taueff$_{V}) - 
\log_{10}(Z/Z_\odot)$(\taueff$_{V}$ = 0).  Reliable
fits are indicated with a check mark, \checkmark.  For unreliable fits the
check mark is replaced with a code indicating the reason for which they
could not be fit.  The codes are: ``{\small OG}'' if the point lies off the 
model grids, ``{\small DB}'' if it lies in the young double-valued Balmer 
line region, and ``{\small DM}'' if it lies in a degenerate metallicity 
region (only occurred for the young exponential SFH model in the 
\dnf\ -- Fe4668 plane).  Note that, in the absence of a reliable fit for the
dust-free case, even if reliable fits exist for the corresponding dusty 
models, the $\Delta$'s are not well defined (only occurred for the 0.5 Gyr
exponential SFH model in Table~\ref{tab:errs_EXP} in the \dnf\ -- Fe4668 
plane).  Examples of the messy double-valued Balmer line region can be seen in
the H$\beta$ versus Mg{\it b} plot in Figure~\ref{fig:indexgrids} [left 
panel] where the ages plotted extend down to 0.002 Gyr,
as well as in Figures~\ref{fig:Hb_Fe} \& \ref{fig:HdelA_d4000} in 
\S\ref{sec:disc}.

It is obvious from Tables~\ref{tab:errs_SSP} \& \ref{tab:errs_EXP} 
(and see Fig.~\ref{fig:montecarlo}) that there is significant uncertainty in 
the fits given our assumed ``typical'' measurement errors, regardless of the 
dust content.  For example, the 0.5 Gyr SSP model 1$\sigma$ confidence limits 
(Table~\ref{tab:errs_SSP}) range from $\sim$0.1--0.4 Gyr in age and 
$\sim$0.2--0.3 dex in metallicity, depending on the index--index grid used. 
Note that the 1$\sigma$ confidence limits are not only dependent on the
particular index-index plane used, but they also depend on the location 
of the point within the grid (in other words, to what degree the age and 
metallicity are separated at the given locus).  In general, the degeneracies 
become more severe at the extreme 
metallicities (\ie\ $Z \la 0.001$ and $Z \ga 0.02$), and at young ages, where 
both the narrower region covered by the metallicity span and the double-valued
Balmer lines contribute to the problem.  

As expected from \S\ref{sec:results}, the fits for the 0.5 Gyr SSP models 
(Table~\ref{tab:errs_SSP}) are virtually 
unaffected by dust (for the index--index planes shown), with the exception of 
the \dnf\ -- Fe4668 plane, for which the age and metallicity errors are large 
(up to 9.5 Gyr and 0.8 dex 
respectively) and significant compared to the 1$\sigma$ confidence limits,
even for small amounts of dust. Significant errors occur at the highest dust 
opacities for the 5 Gyr models (in all but the H$\beta$ -- Mg{\it b} and
[MgFe]$\arcmin$ planes), but in this case they are of the same 
order as the 1$\sigma$ confidence limits.  The errors due to dust on age and 
metallicity are most severe in the 13 Gyr models, with some points ending up 
off the model grids at the highest \taueff$_{V}$'s (\eg\ in the H$\beta$ 
and H$\delta_{A}$ versus Fe4668 grids), but these are again matched by the 
large 1$\sigma$ confidence limits for fits in these regions of the grid.
In summary, for the SSP model galaxies, if the model indices lie in a region
of the index--index diagnostic plot where the age and metallicity are well
separated (close to orthogonal), the fits are generally independant of dust 
extinction for most index combinations.  Where the fits are affected by the 
dust (generally at older ages and for indices with broader baselines), the
errors in the derived ages and metallicities are generally of the same order
as the 1$\sigma$ confidence limits and would thus be difficult to detect.
The most notable exceptions here are the H$\beta$--Mg$_{2}$ plane, where the 
dust pushes the points for the 0.5 Gyr models into the degenerate 
double-valued Balmer line region, and the \dnf\ -- Fe4668 plane, where the 
errors due to dust become large at young ages, and at old ages the points 
extend off the edges of the model grids.  

On the other hand, for the exponential SFH models, again as expected from 
\S\ref{sec:results}, the situation is more complicated.
Examination of Table~\ref{tab:errs_EXP} reveals that none of the 0.5 Gyr models
can be fit reliably in the Balmer-index versus metallicity-index grids
as they all fall in the double-valued Balmer line 
region of the plots.  Reliable fits for the dustier 0.5 Gyr models in
the \dnf\ -- Fe4668 plane are obtained, but here the $\Delta$'s are undefined 
due to the poor fit in the dust-free model.  Reliable fits are
found for most of the 5 Gyr exponential SFH models with ``SSP ages'' in the 
range $\sim$0.9--1.0 Gyr (\ie\ this is the luminosity-weighted SSP age of a 
5 Gyr old stellar population that has been forming stars at a constant rate, 
roughly, throughout its lifetime).  The ages, when reliably fit, generally 
agree between the different index--index plane fits.  When dust has an 
effect on the ages, it tends to make them slightly youger, but the $\Delta$A's 
are always of the same order as the 1$\sigma$ confidence limits for the fits.  
The metallicity fits for these models, which are all inherently solar 
metallicity, are in the range $\log_{10}(Z/Z_\odot) \sim -0.7$ to $+0.2$
(or $Z$ = 0.004 -- 0.03).  They are generally lower(higher) than solar in the 
5 (13) Gyr models and increase
with dust extinction.  Unlike the age fits, the metallicity fits do not 
agree very well between the different index--index planes.  For example,
the \taueff$_{V}=0$ fit in 13 Gyr models gives $\log_{10}(Z/Z_\odot)$ = +0.32 
and $-$0.12 dex in the H$\beta$ -- Mg{\it b} and  H$\beta$ -- Fe4668 planes 
respectively.  The ages for the 13 Gyr exponential SFH model fits are also in 
the $\sim$0.9--1.0 Gyr range.  While these do have weaker Balmer lines than the
5 Gyr models, their metallicity index values are slightly stronger.  This
has the overall effect of moving them along a line of roughly constant age.
An example of this is shown by the 5 Gyr (green) and 13 Gyr (blue) triangles 
in the H$\beta$ versus Mg{\it b} diagnostic plot in 
Figure~\ref{fig:indexgrids} [left panel].  These results highlight once again 
the greater challenge of fitting reliable ages and metallicities for ``young'' 
populations.

\end{appendix}

\clearpage

\renewcommand{\baselinestretch}{1.3}

\begin{landscape}
\begin{deluxetable}{c|c|c|rrrr|c|rrrr|c|rrrr}
\tablewidth{0pt}
\tabletypesize{\scriptsize}
\tablecaption{Age \& Metallicity Fits and Errors: SSP with $Z=Z_{\odot}$ and $
\mu$ = 1.0. 
\label{tab:errs_SSP}}
\tablehead{
\\ [-15pt]
\multicolumn{1}{c|}{} & 
\multicolumn{1}{c|}{} & 
\multicolumn{5}{c|}{H$\beta$ vs. Mg{\it b}} & 
\multicolumn{5}{c|}{H$\beta$ vs. Mg$_2$} & 
\multicolumn{5}{c}{H$\beta$ vs. [MgFe]$\arcmin$} \\ [1pt]
\hline
\\ [-8pt]
\multicolumn{1}{c|}{{\tiny MA}} & 
\multicolumn{1}{c|}{\taueff$_{V}$} & 
\multicolumn{1}{c|}{fit\tablenotemark{a}} & 
\colhead{Age} & 
\colhead{$\Delta$A} & 
\colhead{log($Z/Z_{\odot}$)} & 
\multicolumn{1}{c|}{$\Delta Z$} &
\multicolumn{1}{c|}{fit\tablenotemark{a}} & 
\colhead{Age} & 
\colhead{$\Delta$A} & 
\colhead{log($Z/Z_{\odot}$)} & 
\multicolumn{1}{c|}{$\Delta Z$} &
\multicolumn{1}{c|}{fit\tablenotemark{a}} & 
\multicolumn{1}{c}{Age} & 
\colhead{$\Delta$A} & 
\colhead{log($Z/Z_{\odot}$)} & 
\colhead{$\Delta Z$}}
\startdata
\\[-16pt]
 0.5 & 0 & \checkmark  &  0.51 (0.22) &  0.00 &  0.00  (0.35) &  0.00 &   {\tiny DB}        &  0.51  (0.43) &  0.00 &  0.00  (0.38) &  0.00 & \checkmark  &  0.51  (0.10) &  0.00 &  0.00  (0.33) &  0.00  \\
    & 2 & \checkmark  &  0.51  (0.22) &  0.00 &  0.00  (0.36) &  0.00 &   {\tiny DB}        &  0.45  (0.36) & -0.06 & -0.44  (0.74) & -0.44 & \checkmark  &  0.51  (0.10) &  0.00 & -0.01  (0.33) & -0.01  \\
    & 4 & \checkmark  &  0.51  (0.22) &  0.00 &  0.00  (0.36) &  0.00 &   {\tiny DB}        &  0.45  (0.36) & -0.06 & -0.46  (0.77) & -0.46 & \checkmark  &  0.51  (0.10) &  0.00 & -0.01  (0.32) & -0.01  \\
    & 6 & \checkmark  &  0.51  (0.22) &  0.00 &  0.00  (0.35) &  0.00 &   {\tiny DB}        &  0.09  (0.19) & -0.42 &  0.29  (0.38) &  0.29 & \checkmark  &  0.51  (0.10) &  0.00 &  0.00  (0.34) &  0.00  \\
 
\hline
 5  & 0 & \checkmark  &  5.00  (1.63) &  0.00 &  0.00  (0.13) &  0.00 & \checkmark  &  5.00  (1.50) &  0.00 &  0.00  (0.09) &  0.00 & \checkmark  &  5.00  (1.63) &  0.00 &  0.00  (0.13) &  0.00  \\
    & 2 & \checkmark  &  5.00  (1.50) &  0.00 &  0.00  (0.12) &  0.00 & \checkmark  &  5.00  (1.50) &  0.00 &  0.00  (0.09) &  0.00 & \checkmark  &  5.00  (1.50) &  0.00 &  0.00  (0.12) &  0.00  \\
    & 4 & \checkmark  &  5.00  (1.50) &  0.00 &  0.00  (0.12) &  0.00 & \checkmark  &  4.50  (1.63) & -0.50 &  0.06  (0.13) &  0.06 & \checkmark  &  5.00  (1.75) &  0.00 &  0.00  (0.14) &  0.00  \\
    & 6 & \checkmark  &  5.00  (1.50) &  0.00 &  0.00  (0.12) &  0.00 & \checkmark  &  4.25  (1.57) & -0.75 &  0.11  (0.18) &  0.11 & \checkmark  &  4.75  (1.75) & -0.25 &  0.04  (0.14) &  0.04  \\
 
\hline
 13 & 0 & \checkmark  & 13.00  (6.25) &  0.00 &  0.00  (0.21) &  0.00 & \checkmark  & 13.00  (4.38) &  0.00 &  0.00  (0.13) &  0.00 & \checkmark  & 13.00  (5.88) &  0.00 &  0.00  (0.22) &  0.00  \\
    & 2 & \checkmark  & 11.50  (6.00) & -1.50 &  0.08  (0.20) &  0.08 & \checkmark  & 13.25  (4.38) &  0.25 &  0.00  (0.13) &  0.00 & \checkmark  & 12.25  (6.00) & -0.75 &  0.04  (0.22) &  0.04  \\
    & 4 & \checkmark  & 10.50  (6.25) & -2.50 &  0.13  (0.21) &  0.13 & \checkmark  & 11.50  (4.75) & -1.50 &  0.08  (0.13) &  0.08 & \checkmark  & 11.25  (6.00) & -1.75 &  0.10  (0.22) &  0.10  \\
    & 6 & \checkmark  & 10.50  (5.87) & -2.50 &  0.13  (0.20) &  0.13 & \checkmark  &  9.50  (4.75) & -3.50 &  0.19  (0.15) &  0.19 & \checkmark  & 10.00  (6.00) & -3.00 &  0.16  (0.22) &  0.16  \\ 
\hline
\hline
\multicolumn{2}{c}{} & 
\multicolumn{5}{c|}{H$\beta$ vs. Fe4668} &
\multicolumn{5}{c|}{H$\delta_A$ vs. Fe4668} &
\multicolumn{5}{c}{\dnf\ vs. Fe4668} \\
\hline
\hline
 0.5 & 0 & \checkmark  & 0.51  (0.10) &  0.00 &  0.00  (0.31) &  0.00 & \checkmark  &  0.51  (0.10) &  0.00 &  0.00  (0.18) &  0.00 & \checkmark  &  0.51  (0.12) &  0.00 &  0.00  (0.34) &  0.00  \\
    & 2 & \checkmark  &  0.51  (0.10) &  0.00 & -0.03  (0.34) & -0.03 & \checkmark  &  0.51  (0.15) &  0.00 & -0.02  (0.15) & -0.02 & \checkmark  &  0.90  (0.48) &  0.40 & -0.44  (0.27) & -0.44  \\
    & 4 & \checkmark  &  0.51  (0.10) &  0.00 & -0.03  (0.34) & -0.03 & \checkmark  &  0.51  (0.10) &  0.00 & -0.02  (0.16) & -0.02 & \checkmark  &  2.60  (2.80) &  2.09 & -0.74  (0.10) & -0.74  \\
    & 6 & \checkmark  &  0.51  (0.10) &  0.00 &  0.00  (0.31) &  0.00 & \checkmark  &  0.51  (0.06) &  0.00 &  0.00  (0.09) &  0.00 & \checkmark  & 10.00  (3.13) &  9.49 & -0.80  (0.05) & -0.80  \\
 
\hline
 5  & 0 & \checkmark  &  5.00  (1.93) &  0.00 &  0.00  (0.22) &  0.00 & \checkmark  &  5.00  (2.33) &  0.00 &  0.00  (0.22) &  0.00 & \checkmark  &  5.00  (2.03) &  0.00 &  0.00  (0.23) &  0.00  \\
    & 2 & \checkmark  &  5.00  (1.80) &  0.00 &  0.00  (0.22) &  0.00 & \checkmark  &  5.00  (2.38) &  0.00 &  0.00  (0.23) &  0.00 & \checkmark  & 11.00  (2.12) &  6.00 & -0.12  (0.05) & -0.12  \\
    & 4 & \checkmark  &  3.50  (1.93) & -1.50 &  0.18  (0.22) &  0.18 & \checkmark  &  4.00  (2.05) & -1.00 &  0.15  (0.23) &  0.15 & {\tiny OG}  & 20.00  (2.00) & 15.00 & -0.15  (0.02) & -0.15  \\
    & 6 & \checkmark  &  3.00  (1.80) & -2.00 &  0.29  (0.22) &  0.29 & \checkmark  &  3.25  (1.93) & -1.75 &  0.24  (0.22) &  0.24 & {\tiny OG}  & 20.00  (0.00) & 15.00 & -0.07  (0.02) & -0.07  \\
 
\hline
 13 & 0 & \checkmark  & 13.00  (3.00) &  0.00 &  0.00  (0.12) &  0.00 & \checkmark  & 13.00  (4.00) &  0.00 &  0.00  (0.20) &  0.00 & \checkmark  & 13.00  (2.50) &  0.00 &  0.00  (0.11) &  0.00  \\
    & 2 & \checkmark  & 12.75  (3.12) & -0.25 &  0.02  (0.11) &  0.02 & \checkmark  & 12.75  (4.13) & -0.25 &  0.02  (0.21) &  0.02 & {\tiny OG}  & 20.00  (1.50) &  7.00 & -0.02  (0.04) & -0.02  \\
    & 4 & {\tiny OG}  & 12.00  (3.12) & -1.00 &  0.06  (0.10) &  0.06 & \checkmark  &  6.25  (8.25) & -6.75 &  0.39  (0.42) &  0.39 & {\tiny OG}  & 17.00  (0.00) &  4.00 &  0.37  (0.03) &  0.37  \\
    & 6 & {\tiny OG}  & 12.25  (2.87) & -0.75 &  0.06  (0.09) &  0.06 &  {\tiny OG} &  7.50  (3.38) & -5.50 &  0.37  (0.21) &  0.37 & {\tiny OG}  & 20.00  (0.00) &  7.00 &  0.40  (0.00) &  0.40 \\ [-11pt]
\enddata
\tablecomments{All ages are given in Gyr and metallicities as 
$\log_{10}(Z/Z_{\odot})$.  The numbers in brackets next to the age and 
metallicity measurements are the 1$\sigma$ confidence limits on
the derived values based on 1000 Monte Carlo realizations drawn from a 
Gaussian distribution of the measurement errors.}
\tablenotetext{a}{A check mark, \checkmark, indicates a reliable fit.  
Unreliable fits are coded as: ``{\tiny OG}'' if the point lies off the 
model grids, and ``{\tiny DB}'' if it lies in the young double-valued Balmer 
line region.}
\end{deluxetable}
\clearpage
\end{landscape}

\clearpage

\begin{landscape}
\begin{deluxetable}{c|c|c|rrrr|c|rrrr|c|rrrr}
\tablewidth{0pt}
\tabletypesize{\scriptsize}
\tablecaption{Age \& Metallicity Errors: exponential SFH with $\tau_{exp} = 13$ Gyr and $\mu$ = 0.3. 
\label{tab:errs_EXP}}
\tablehead{
\\ [-15pt]
\multicolumn{1}{c|}{} & 
\multicolumn{1}{c|}{} & 
\multicolumn{5}{c|}{H$\beta$ vs. Mg{\it b}} & 
\multicolumn{5}{c|}{H$\beta$ vs. Mg$_2$} & 
\multicolumn{5}{c}{H$\beta$ vs. [MgFe]$\arcmin$} \\ [1pt]
\hline  
\\ [-8pt]
\multicolumn{1}{c|}{{\tiny MA}} & 
\multicolumn{1}{c|}{\taueff$_{V}$} & 
\multicolumn{1}{c|}{fit\tablenotemark{a}} & 
\colhead{Age} & 
\colhead{$\Delta$A} & 
\colhead{log($Z/Z_{\odot}$)} & 
\multicolumn{1}{c|}{$\Delta Z$} &
\multicolumn{1}{c|}{fit\tablenotemark{a}} & 
\colhead{Age} & 
\colhead{$\Delta$A} & 
\colhead{log($Z/Z_{\odot}$)} & 
\multicolumn{1}{c|}{$\Delta Z$} &
\multicolumn{1}{c|}{fit\tablenotemark{a}} & 
\multicolumn{1}{c}{Age} & 
\colhead{$\Delta$A} & 
\colhead{log($Z/Z_{\odot}$)} & 
\colhead{$\Delta Z$}}
\startdata
\\[-16pt]
 0.5 & 0 & {\tiny DB} &  0.08  (0.02) &  0.00 & -0.49  (0.50) &  0.00 &  {\tiny DB} &  0.06  (0.03) &  0.00 & -0.10  (0.52) &  0.00 &  {\tiny DB} &  0.06  (0.33) &  0.00 & -0.01  (0.43) &  0.00  \\
    & 2 &  {\tiny DB} &  0.11  (0.04) &  0.03 & -0.60  (0.21) & -0.11 &  {\tiny DB} &  0.64  (0.60) &  0.58 & -0.60  (0.53) & -0.51 &  {\tiny DB} &  0.64  (0.54) &  0.58 & -0.82  (0.19) & -0.82  \\
    & 4 &  {\tiny DB} &  0.14  (0.26) &  0.06 & -0.62  (0.21) & -0.12 &  {\tiny DB} &  0.08  (0.27) &  0.02 &  0.02  (0.53) &  0.12 &  {\tiny DB} &  0.57  (0.19) &  0.51 & -0.76  (0.10) & -0.75  \\
    & 6 &  {\tiny DB} &  0.14  (0.26) &  0.06 & -0.60  (0.22) & -0.11 &  {\tiny DB} &  0.03  (0.54) & -0.03 &  0.40  (1.08) &  0.49 &  {\tiny DB} &  0.57  (0.19) &  0.51 & -0.76  (0.08) & -0.75  \\
    & 8 &  {\tiny DB} &  0.14  (0.22) &  0.06 & -0.60  (0.19) & -0.11 &  {\tiny DB} &  0.09  (0.26) &  0.03 & -0.05  (0.51) &  0.05 &  {\tiny DB} &  0.26  (0.39) &  0.20 & -0.70  (0.09) & -0.69  \\
 
\hline
 5  & 0 & \checkmark  &  1.02  (0.11) &  0.00 & -0.18  (0.45) &  0.00 & \checkmark  &  0.90  (0.10) &  0.00 &  0.10  (0.19) &  0.00 & \checkmark  &  1.02  (0.00) &  0.00 & -0.22  (0.19) &  0.00  \\
    & 2 & \checkmark  &  1.02  (0.21) &  0.00 & -0.68  (0.90) & -0.50 & \checkmark  &  0.81  (0.10) & -0.10 &  0.22  (0.12) &  0.12 & \checkmark  &  0.90  (0.11) & -0.11 &  0.04  (0.22) &  0.26  \\
    & 4 & \checkmark  &  0.81  (0.21) & -0.21 &  0.20  (0.41) &  0.38 & \checkmark  &  0.81  (0.10) & -0.10 &  0.20  (0.12) &  0.11 & \checkmark  &  0.90  (0.10) & -0.11 &  0.04  (0.22) &  0.26  \\
    & 6 & \checkmark  &  0.81  (0.21) & -0.21 &  0.20  (0.36) &  0.38 & \checkmark  &  0.81  (0.09) & -0.10 &  0.20  (0.12) &  0.11 & \checkmark  &  0.90  (0.10) & -0.11 &  0.04  (0.21) &  0.26  \\
    & 8 & \checkmark  &  0.81  (0.21) & -0.21 &  0.20  (0.34) &  0.38 & \checkmark  &  0.81  (0.10) & -0.10 &  0.20  (0.12) &  0.11 & \checkmark  &  0.90  (0.10) & -0.11 &  0.04  (0.22) &  0.26  \\
 
\hline
 13 & 0 & \checkmark  &  0.90  (0.11) &  0.00 &  0.32  (0.07) &  0.00 & \checkmark  &  0.90  (0.11) &  0.00 &  0.33  (0.08) &  0.00 & \checkmark  &  1.02  (0.12) &  0.00 &  0.20  (0.18) &  0.00  \\
    & 2 & \checkmark  &  0.90  (0.10) &  0.00 &  0.33  (0.04) &  0.01 & \checkmark  &  0.90  (0.10) &  0.00 &  0.33  (0.04) &  0.00 & \checkmark  &  0.90  (0.11) & -0.11 &  0.32  (0.07) &  0.12  \\
    & 4 & \checkmark  &  0.90  (0.10) &  0.00 &  0.33  (0.04) &  0.01 & \checkmark  &  0.90  (0.10) &  0.00 &  0.33  (0.04) &  0.00 & \checkmark  &  0.90  (0.10) & -0.11 &  0.32  (0.07) &  0.12  \\
    & 6 & \checkmark  &  0.90  (0.10) &  0.00 &  0.33  (0.04) &  0.01 & \checkmark  &  0.90  (0.10) &  0.00 &  0.33  (0.04) &  0.00 & \checkmark  &  0.90  (0.10) & -0.11 &  0.32  (0.06) &  0.12  \\
    & 8 & \checkmark  &  0.90  (0.10) &  0.00 &  0.33  (0.04) &  0.01 & \checkmark  &  0.90  (0.10) &  0.00 &  0.33  (0.04) &  0.00 & \checkmark  &  0.90  (0.10) & -0.11 &  0.32  (0.06) &  0.12  \\
 
\hline  
\hline 
\multicolumn{1}{c}{} & 
\multicolumn{2}{c|}{} & 
\multicolumn{4}{c|}{H$\beta$ vs. Fe4668} & 
\multicolumn{5}{c|}{H$\delta_{A}$ vs. Fe4668} & 
\multicolumn{5}{c}{\dnf\ vs. Fe4668} \\
\hline  
\hline  

 0.5 & 0 & {\tiny DB}  &  0.04  (0.34) &  0.00 &  0.22  (0.44) &  0.00 & {\tiny DB}  &  0.04  (0.98) &  0.00 &  0.28  (0.99) &  0.00 &  {\tiny DM} &  0.07  (0.05) &  0.00 & -0.01  (0.32) &  0.00  \\
    & 2 &  {\tiny DB}  &  0.64  (0.58) &  0.60 & -0.60  (0.72) & -0.82 & {\tiny DB}  &  0.81  (0.42) &  0.77 & -0.72  (0.76) & -1.00 &  {\tiny OG} &  0.18  (0.11) &  0.11 &  0.40  (0.80) &  0.41  \\
    & 4 &  {\tiny DB}  &  0.64  (0.57) &  0.60 & -0.54  (0.33) & -0.76 & {\tiny DB}  &  0.81  (0.72) &  0.77 & -0.72  (0.74) & -1.00 & \checkmark  &  0.20  (0.04) &  0.13 &  0.37  (0.32) &  0.38  \\
    & 6 &  {\tiny DB}  &  0.64  (0.57) &  0.60 & -0.54  (0.32) & -0.76 & {\tiny DB}  &  0.81  (0.72) &  0.77 & -0.72  (0.74) & -1.00 & \checkmark  &  0.26  (0.05) &  0.18 &  0.10  (0.39) &  0.11  \\
    & 8 &  {\tiny DB}  &  0.64  (0.57) &  0.60 & -0.54  (0.32) & -0.76 & {\tiny DB}  &  0.81  (0.72) &  0.77 & -0.72  (0.72) & -1.00 & \checkmark  &  0.32  (0.07) &  0.25 & -0.22  (0.34) & -0.21  \\
 
\hline
 5  & 0 & \checkmark  &  1.02  (0.00) &  0.00 & -0.24  (0.07) &  0.00 & \checkmark  &  1.02  (0.12) &  0.00 & -0.25  (0.14) &  0.00 &  {\tiny OG} &  0.32  (0.03) &  0.00 &  0.40  (0.00) &  0.00  \\
    & 2 & \checkmark  &  1.02  (0.11) &  0.00 & -0.24  (0.12) &  0.00 & \checkmark  &  0.90  (0.10) & -0.11 & -0.09  (0.16) &  0.17 &  {\tiny OG} &  0.45  (0.05) &  0.13 &  0.39  (0.01) & -0.01  \\
    & 4 & \checkmark  &  1.02  (0.11) &  0.00 & -0.24  (0.13) &  0.00 & \checkmark  &  0.90  (0.15) & -0.11 & -0.09  (0.24) &  0.17 &  {\tiny OG} &  0.45  (0.05) &  0.13 &  0.40  (0.00) &  0.00  \\
    & 6 & \checkmark  &  1.02  (0.11) &  0.00 & -0.24  (0.14) &  0.00 & \checkmark  &  0.90  (0.15) & -0.11 & -0.09  (0.23) &  0.17 &  {\tiny OG} &  0.51  (0.06) &  0.19 &  0.38  (0.02) & -0.02  \\
    & 8 & \checkmark  &  1.02  (0.11) &  0.00 & -0.24  (0.12) &  0.00 & \checkmark  &  0.90  (0.15) & -0.11 & -0.09  (0.24) &  0.17 &  {\tiny OG} &  0.51  (0.06) &  0.19 &  0.40  (0.05) &  0.00  \\
 
\hline
 13 & 0 & \checkmark  &  1.14  (0.12) &  0.00 & -0.12  (0.20) &  0.00 & \checkmark  &  1.02  (0.17) &  0.00 &  0.06  (0.26) &  0.00 & {\tiny OG}  &  0.57  (0.07) &  0.00 &  0.40  (0.00) &  0.00  \\
    & 2 & \checkmark  &  1.02  (0.12) & -0.12 &  0.18  (0.20) &  0.30 &  {\tiny OG} &  0.81  (0.33) & -0.21 &  0.37  (0.22) &  0.31 & {\tiny OG}  &  1.02  (0.37) &  0.44 &  0.00  (0.40) & -0.40  \\
    & 4 & \checkmark  &  1.02  (0.12) & -0.12 &  0.18  (0.19) &  0.30 &  {\tiny OG} &  0.81  (0.33) & -0.21 &  0.37  (0.21) &  0.31 & {\tiny OG}  &  1.02  (0.28) &  0.44 &  0.00  (0.34) & -0.40  \\
    & 6 & \checkmark  &  1.02  (0.12) & -0.12 &  0.18  (0.19) &  0.30 &  {\tiny OG} &  0.81  (0.33) & -0.21 &  0.37  (0.21) &  0.31 & {\tiny OG}  &  1.02  (0.26) &  0.44 &  0.00  (0.00) & -0.40  \\
    & 8 & \checkmark  &  1.02  (0.12) & -0.12 &  0.18  (0.19) &  0.30 &  {\tiny OG} &  0.81  (0.33) & -0.21 &  0.37  (0.21) &  0.31 & {\tiny OG}  &  1.28  (0.26) &  0.71 &  0.00  (0.02) & -0.40  \\ [-11pt]
\enddata
\tablecomments{All ages are given in Gyr and metallicities as 
$\log_{10}(Z/Z_{\odot})$.  The numbers in brackets next to the age and 
metallicity measurements are the 1$\sigma$ confidence limits on
the derived values based on 1000 Monte Carlo realizations drawn from a 
Gaussian distribution of the measurement errors.  
Note that the $\Delta$'s are not well defined for the 0.5 Gyr models in 
the \dnf\ vs. Fe4668 that are marked as reliable fits, since the 
dust-free case cannot be reliably fit.}
\tablenotetext{a}{
A check mark, \checkmark, indicates a reliable fit.  
Unreliable fits are coded as: ``{\tiny OG}'' if the point lies off the 
model grids, ``{\tiny DB}'' if it lies in the young double-valued Balmer 
line region, and ``{\tiny DM}'' if it lies in a degenerate metallicity 
region (only occurred for the young exponential SFH model in the 
\dnf\ -- Fe4668 plane).}
\end{deluxetable}
\clearpage
\end{landscape}
\renewcommand{\baselinestretch}{1.0}

\end{document}